\documentclass[11pt,twocolumn]{article}
\usepackage{authblk}
\usepackage{amsmath,amssymb}
\usepackage{hyperref}

\usepackage[utf8]{inputenc}

\usepackage{amsmath} \usepackage{xspace}

\usepackage{amsmath,amsfonts,textcomp} 
\usepackage{mathtools}
\usepackage{mathrsfs}
\usepackage{graphicx} \graphicspath{{figs/}} \usepackage{xcolor}

\graphicspath{{figs/}}

\makeatletter \newcommand{\MathFuncName}[1]{{\operator@font #1}}
\newcommand{\MathFunc}[1]{\mathop{\operator@font #1}\nolimits}
\newcommand{\MathFuncWithLimits}[1]{\mathop{\operator@font #1}\limits}
\makeatother

\newcommand{\changes}[1]{#1}

\newcommand{\mathd}{\mathrm{d}} 
\newcommand{\mathe}{\mathrm{e}} 

\newcommand{\RM}[1]{\mathrm{#1}} 
\newcommand{\V}[1]{\boldsymbol{#1}} 
\newcommand{\M}[1]{\mathbf{#1}} 
\newcommand{\AdjointLetter}{\top}
\newcommand{\T}{^{\AdjointLetter}} 
\newcommand{\TransposeLetter}{\top}
\newcommand{\Tt}{^{\TransposeLetter}} 
\newcommand{\E}{\mathe} 
\newcommand{\D}[1]{{\mathd #1}} 
\newcommand{\sign}{\MathFunc{sign}} 
\newcommand{\Arg}{\MathFunc{Arg}} 
\newcommand{\comp}{\mathop{\circ}}

\newcommand{\Eq}[1]{Eq.~(\ref{#1})}
\newcommand{\Fig}[1]{Fig.~\ref{#1}}
\newcommand{\Sec}[1]{Sec.~\ref{#1}}

\newcommand{\YORICK}[1]{}

\newcommand{\Paren}[1]{\left(#1\right)}

\newcommand{\Norm}[1]{\left\Vert #1\right\Vert}

\newcommand{\abs}[1]{\vert #1\vert} 
\newcommand{\Abs}[1]{\left\vert #1\right\vert}

\newcommand{\Avg}[1]{\left\langle #1\right\rangle}

\newcommand{\FT}[1]{\widehat{#1}} 
\newcommand{\F}[1]{\mathcal{F}\Paren{#1}}
\newcommand{\IF}[1]{\mathcal{F}^{-1}\Paren{#1}}

 \newcommand{\Reals}{\mathbb{R}}
 
\newcommand{\Complexes}{\mathbb{C}}

\newcommand{\eg}{\emph{e.g.}\xspace}
\newcommand{\ie}{\emph{i.e.}\xspace} \newcommand{\etal}{\emph{et
    al.}\xspace}

\newcommand{\micron}{\text{\textmu{}m}} 
\newcommand{\nm}{\text{nm}} 
\newcommand{\NA}{\textrm{NA}}
\newcommand{\fov}{FoV\xspace}
\newcommand{\sbp}{SBP\xspace}

\begin{document}

\title{Gauging diffraction patterns: field of view and bandwidth
  estimation in lensless holography}

\author{Ferréol Soulez\footnote{ferreol.soulez@univ-lyon1.fr}}

\affil{ Univ.\ Lyon, Univ.\ Lyon 1, ENS de Lyon, CNRS, Centre de
  Recherche Astrophysique de Lyon UMR5574, F-69230, Saint-Genis-Laval,
  France}



\maketitle
\begin{abstract} 
The purpose of this work is to provide theoretically grounded assessment  on both the field-of-view and the bandwidth of a lensless holographic setup. Indeed, while previous works have presented results with super-resolution and field-of-view extrapolation, there is no well established rules to determine them. We show that the theoretical field of view can be hugely large with a \changes{spatial-frequency} bandwidth only limited by the wavelength leading to an unthinkable number of degrees of freedom.  To keep a realistic field of view and bandwidth, we propose  several practical bounds based on few setup properties: namely the noise level and the spatio-temporal coherence of the source.
\end{abstract}


\section{Introduction}
Lensless in-line holography   consists in directly recording the light diffracted by
the observed sample on a detector without any optical parts between
them. Given the simplicity, the compactness, the robustness  and the 
relatively low cost of this setup \cite{Mudanyali2010,Rostykus2017,Rostykus2018}, inline digital holography is successfully employed in many applications such as lensfree microscopy\cite{Rostykus2017,Rostykus2018,Allier2010,Mudanyali2013} or  metrology\cite{Murata2000}.

Contrary to direct imaging methods,  the recorded hologram cannot be directly interpreted  and
computational algorithms are mandatory to recover an image of the
sample or to extract any parameters of interest. This reconstruction
step has its own limitations and it is difficult to disentangle  whether the effective resolution and the field of view (\fov) of the reconstructed image are due to some physical limits of the setup or to an imperfect reconstruction algorithm:
\begin{itemize}
    \item Concerning the \fov,  many works     \cite{Schnars1994,GoodmanFourier,Milgram2002,Dubois2002,Migukin2011,Latychevskaia2013,Bishara2010,Zhang2017,Luo2019} restrict the lateral \fov{} of the reconstructed image to the \fov{} of the sensor. However  for similar measurements, some reconstruction methods \cite{Kreis1997,Soulez2007,Denis2009,Fournier2016} achieve to recover information on a \fov{} much larger than that of the sensor (detection in an area 16 times wider than camera \fov{} in the case of \cite{Soulez2007}). Nonetheless none of these works provide any estimation of the size of the extrapolated \fov. 
    \item Without any lens to filter it, the wave-field in the detector plane may contain angular spatial frequencies as high as the  wavenumber of illumination $k$. However, many works \cite{Schnars1994,GoodmanFourier,Milgram2002,Dubois2002,Migukin2011,Latychevskaia2013} reconstruct the object at the sampling rate given by the detector pixel pitch that can be order of magnitude coarser. To overcome this limitation, several methods were proposed to  recover aliased spatial frequencies either using super-resolution methods\cite{Greenbaum2013,Fournier2016,Zhang2017,Luo2015,Bishara2010,Luo2019} or using prior knowledge  on the observed scene \cite{Onural2000,Stern2006,Soulez2007a,Fournier2010}.
\end{itemize}
A review of the state of the art of reconstruction methods shows no consensus about how to assess the resolution and the \fov{} of a given setup. The purpose of this work is thus to answer the question: \textit{What are the effective field of view and the \changes{spatial-frequency} bandwidth of a lensless setup?} 


In the literature, the Rayleigh criterion is the most  commonly used definition of the resolution. In this work, the spatial resolution $R$ is defined as the inverse of the highest spatial frequency transmitted by the setup. For a setup of angular bandwidth $B$, the resolution is $R=  {4\,\pi}/{B}$. 
The resolution limits of an holographic system has been discussed by many authors \cite{Stern2006,Kelly2009,Fournier2010,Hao2011,Agbana2017,Zhang2020}. Contrary to what is presented here, most of these works consider the resolution of the whole system (propagation + sensing + reconstruction) and aliasing issues. In this paper, we propose bounds on the extent in both space and frequency of the sample to model the measured intensity with the highest fidelity.  
This gauging of the bandwidth and the \fov of a setup does not say anything about how  the propagation kernel should be  numerically implemented and how the object should be reconstructed in further numerical processing steps or even if it is possible. Precise and numerically efficient propagation modeling is a research subject in itself and has been extensively studied either for reconstruction \cite{Chacko2013,Kelly2014,Liu2012,Matsushima2009,Matsushima2010,Odate2011,Onural2000,Ozaktas2011,Falaggis2013,Kozacki2015,Ritter2014,Shimobaba2013,Yu2012} or for hologram generation \cite{Gori1981,Liang2014}.




At optical wavelengths, detectors cannot measure complex amplitude but
only intensity of the light. 
The Fourier spectrum of the
intensity is the auto-correlation of the Fourier spectrum of the complex
amplitude \cite{GoodmanFourier}.  This has two consequences: (i) the
bandwidth of the intensity is twice that of the diffracted wave and
(ii) complex amplitude high spatial frequencies are folded by the auto-correlation and can generate low frequency components in intensity. 
To properly explain the spatial frequency content of the measured intensity, it is necessary to model the complex amplitude at frequencies much higher than the actual sampling rate of the intensity. Hence, the bandwidth  of a lensless inline holography setup  is  independent of the sampling of the intensity by the detector  (sampling rate, pixel shape, pixel fill-factor,\dots).
In a medium of refractive index $n$, only spatial angular frequencies
lower than $k = n\,\frac{2\,\pi}{\lambda}$ can propagate and the angular bandwidth of a perfectly coherent diffracted wave
in the detector plane is  $B = 2\,k$.
Accordingly, regardless of the sampling rate of the camera, to rigorously model the
measured intensity, the diffracted complex amplitude must be sampled with a sampling rate higher  than  $\frac{2\,n}{\lambda}$.



In addition, propagation kernels such as the angular spectrum kernel
are band-limited meaning that they are infinitely extended in the
image domain. As a consequence, the theoretical \fov of an
in-line lensless microscopy is only limited by the size of the
illumination beam. Such large \fov sampled at  $\frac{2\,n}{\lambda}$
lead theoretical space-bandwidth product of several billion pixels
that cannot be handled in practice. The goal of the present paper is to
tighten bounds on both the \fov and the bandwidth to estimate the
actual space-bandwidth product of an experimental lensless holography
setup. 
 To be of  practical interest on real experiments, we focus on giving bounds on the  bandwidth and the \fov only using easily available parameters, namely: illumination angle, coherence length and coherence area of the illumination, size of the sensor and its noise.

\section{Lensless holography model}

\subsection{Setup and notations}

Throughout this paper, we use lower case letters for functions and
scalars (\eg $o$ and $\lambda$), boldface lowercase letters for
vectors (\eg $\V{o}$) upper case calligraphic letters for operators
acting on functions (\eg $\mathcal{M}$) and boldface uppercase letters
for matrices (\eg $\M{H}$).  $\V{x}\T$ is the adjoint (\ie the
conjugate transpose) of $\V{x}$.
$\Norm{\V{x}}_2 = \sqrt{\V{x}\T\V{x}}$ is the Euclidean norm of
$\V{x}$. $\V{x}\T\V{y}$ is the scalar product t between
vectors $\V{x}$ and $\V{y}$ and  $\V{x}\,\times \V{y}$  their element-wise product.

With these notations, each wave is
represented by a square integrable 2D function from
$\Reals^2$ to $\Complexes$ (\eg
$w\, :\, \Reals^2 \rightarrow \Complexes$) and with lateral
coordinates $\V{x}=[x_1,x_2]\Tt $. The discretized version of this wave is the vector
 $\V{w}$ ordered in lexicographical order (\eg $\V{w} = [w_1,\dots,w_N]\Tt$
where $N$ is the number of pixels). Functions and vectors with a hat
(\eg $\FT{w}$) and without a  hat (\eg $w$) are in Fourier domain and
space domain respectively.

\begin{figure}[tbp]
\centering
  \includegraphics[width=0.8\linewidth]{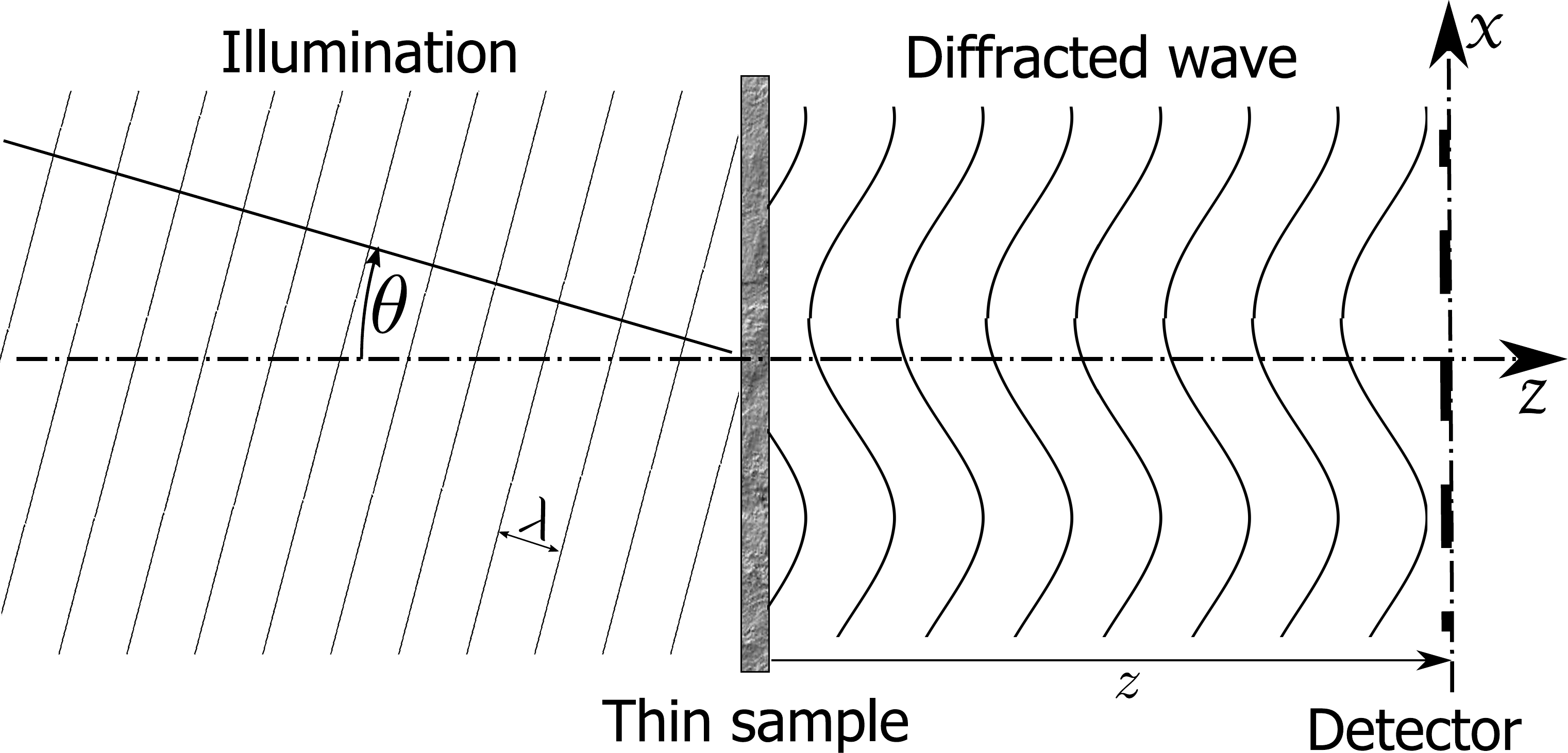}
  \caption{Scheme of the setup}
  \label{fig:Setup}
\end{figure}

"We consider the  lensless setup depicted in \Fig{fig:Setup}:
a thin (2D) sample, described by the function
$o\, :\, \Reals^2 \rightarrow \Complexes$, is placed orthogonally to
the optical axis at $z=0$.  It is illuminated by a plane wave of
wavelength $\lambda$ (or a wavenumber $k = n\frac{2\,\pi}{\lambda}$)
arriving with an incidence angle
$\V{\theta} = [\theta_1, \theta_2]\Tt$ relatively to the optical
axis. After propagation in a medium of refractive index $n$, the
diffracted wave $w$ is recorded by a detector of size
$\ell_1\times\ell_2$ placed at a distance $z$, orthogonally to the
optical axis. The detector produces the discrete  measurements
$\V{d} \in  \Reals^N$.


To give a correct interpretation of the measurements $\V{d}$, one has
to derive a rigorous model accounting for the totality of the measured
information. This is summed up by building the forward operator
$\mathcal{M} :\, L^{2}(\Reals^2) \to \Reals^N$ acting on the Hilbert space of squared integrable functions $L^{2}$ and linking the object
$o$ to the measurements $\V{d}\in  \Reals^N$:
\begin{equation}
  \V{d} = \mathcal{M}(o) + \V{e}\,,
\end{equation}
where $\V{e}$ is an error term.
 
In lensless holography, this forward operator can be described as the
composition of three operators:
\begin{equation}
  \mathcal{M} =  \mathcal{S}\comp \mathcal{C}\comp \mathcal{H}\,,
\end{equation}
where $\comp$ denotes the composition and:\begin{itemize}
\item $ \mathcal{H}:\, L^{2}(\Reals^2) \to L^{2}(\Reals^2)$ models
  the light propagation from the sample plane to the detector
  plane. This linear operator is described in \Sec{sec:propagation}.
\item $ \mathcal{C}:\, L^{2}(\Reals^2) \to L^{2}(\mathbb{D})$ cuts
  the input function on the compact support
  $\mathbb{D} \subset \Reals^2$ describing the sensitive area of the
  detector of size $\V{\ell} = [\ell_1,\ell_2]\Tt$:
  \begin{align}
    \mathcal{C}(f)(\V{x}) &=    \left\{
                            \begin{array}{ll}
                              f(\V{x}) &   \text{if } \V{x}\in\,\mathbb{D} \,, \\
                              0 & \text{otherwise} \,.
                            \end{array}
                                  \right.
  \end{align}
\item $ \mathcal{S}:\, L^{2}(\mathbb{D}) \to \Reals_+^N$ models the
  sensing and the sampling performed by a detector with $N$ pixels. In
  inline holography, the detector samples the intensity of the
  scattered wave.

\end{itemize}

\subsection{Propagation model}
\label{sec:propagation}
 
Right after the interaction of the illumination wave with the sample
$o$, the complex wave field $v$ is:
\begin{equation}
  \label{eq:planewv}
  g(\V{x}) = o(\V{x})
  \exp\left(\jmath \, k\,  \V{x}\T   \sin(\V{\theta})
  \right)  \,,
\end{equation}
where $ \sin(\V{\theta})= [\sin(\theta_1), \sin(\theta_2)]\Tt$ is the component wise 2D sine. 
It can be expressed as a shift in the Fourier domain:
\begin{equation}
  \label{eq:fplanewv}
  \FT{g}(\V{\omega}) =
  \FT{o} \left(\V{\omega}
    -k\,\sin(\V{\theta})\right) \,,
\end{equation}
where $\V{\omega} = [\omega_1,\omega_2]\Tt \in \Reals^2$ is the 2D
angular frequency  and $\mathcal{F}$ is the continuous (non
unitary) 2D Fourier transform operator defined as:
\begin{equation}
  \FT{f}(\V{\omega}) = \F{f} (\V{\omega})= \iint_{\Reals^2} f(\V{x})
  \E^{-\jmath\,\V{x}\T\,\V{\omega}}
  \D{\V{x}}\,.
\end{equation} 
 The property of shifting the sample’s Fourier spectrum by a tilted illumination can be used to perform aperture synthesis as in Fourier ptychography\cite{Zheng2013}. 

Given the high numerical aperture of a lensless setup, the propagation
is modeled in the near-field regime by the mean of the angular
spectrum (AS) propagation mode \cite{GoodmanFourier}. It gives the Fourier transform of the
complex wave field $w$ in the detector plane as:
\begin{equation}
  \FT{w}(\V{\omega}) = \FT{f}^\RM{AS}(\V{\omega}) \,  \FT{g}(\V{\omega}) \,,
\end{equation}
where $\FT{f}^\RM{AS} $ is the angular spectrum transfer function for
$z\gg \lambda$ (neglecting evanescent waves) \cite{GoodmanFourier}:
\begin{equation}
  \label{eq:FAS}
  \FT{f}^\RM{AS}(\V{\omega}) =  \left\{
    \begin{array}{ll}
      \E^{+\jmath\,z\,\sqrt{k^2 - \Norm{\V{\omega}}^2}}\, & \text{if }
                                                            \Norm{\V{\omega}}^2 \le k^2\,\\
      0\, & \text{otherwise.}
    \end{array}\right.\,.
\end{equation}

Let us note here that when most of the propagating wave energy is concentrated on low angular
frequencies, the propagation kernel is non null only when
$\Norm{\V{\omega}}^2 \ll k^2$ leading to the approximation
$\sqrt{k^2 - \Norm{\V{\omega}}^2} \approx k -
\frac{\Norm{\V{\omega}}^2}{2\,k}$. This is the paraxial approximation
and, discarding the constant term $\E^{\jmath\,k\,z}$, the propagation kernel becomes the Fresnel transfer function:
\begin{equation}
  \label{eq:FresnelF}
  \FT{f}^{\RM{F}}(\V{\omega}) =  \E^{-\jmath\,\frac{z}{2\,k}\,\Norm{\V{\omega}}^2}\,.
\end{equation}

The wave field in the detector plane $w$ can be rewritten in the
space domain expressing the operator $\mathcal{H}$ as:
\begin{align}
  w &= \mathcal{H}(o)\,,\\
  w(\V{x}) &=   
             \IF{\FT{h}^\RM{AS} \times \FT{o} } (\V{x})\,
             \E^{\jmath \, k\,  \V{x}\T   \sin(\V{\theta})}   \,,
\end{align}
where $\FT{h}$ is the shifted angular spectrum transfer function
\cite{Guo2014}:
\begin{equation}
  \label{eq:SAS}
  \FT{h}^\RM{AS}(\V{\omega}) =   \left\{
    \begin{array}{ll}
      \E^{\jmath  \,z\,\sqrt{k^2 - \Norm{\V{\omega}
      + k\,\sin(\V{\theta})}^2} } &   \text{if }
                                    \Norm{\V{\omega}+ k\,\sin(\V{\theta})}^2 \le k^2,\\
      0 & \text{otherwise} \,.
    \end{array}
  \right.\,.
\end{equation}
Its expression in space is the Hyugens-Fresnel convolution kernel \cite{GoodmanFourier} under oblique illumination \cite{Guo2014}:
\begin{equation}
  \label{eq:SASS}
  h^\RM{AS}(\V{x}) = \frac{k\, z}{\jmath\, 2 \pi \left(\Norm{\V{x}}^2+ z^2\right)}\E^{\jmath \,k\left(-\V{x}\T\sin(\V{\theta}) + \,\sqrt{\Norm{\V{x}}^2+z^2}\right)}\,.
\end{equation}

\section{Space-bandwidth product analysis}
\label{sec:sbp}
  
The  space-bandwidth product (\sbp) is a powerful tool to assess the performance of optical setups and analyze sampling and reconstruction conditions \cite{Mendlovic1997,Stern2004,Stern2006,Claus2011,Kozacki2015}.
Following Lohmann \etal \cite{Lohmann1996}, we use
the geometrical representation of \sbp in the Wigner domain. For sake of clarity, we consider in this section only 1D signals. In this
representation, the support of a signal of angular bandwidth $B_{S}$
over a \fov $\ell_{s}$ is a rectangle as depicted in
\Fig{fig:SWF}.a

\subsection{Space-bandwidth product in paraxial approximation}
The Wigner distribution function $W^{\RM{F}}$ of the Fresnel transfer
function given \Eq{eq:FresnelF} is \cite{Lohmann1996}:
\begin{align}
  W^{\RM{F}}(x,\omega) &= \int
                         \FT{f}^{\RM{F}}\left(\omega+\frac{\omega'}{2}\right)\,\FT{f}^{\RM{F}*}\left(\omega-\frac{\omega'}{2}\right)
                         \, \E^{\jmath\, \omega'\,x}\,\D{\omega'}\,\\ 
                       &= \delta\left(\omega - x \frac{k}{z}\right)\,,
\end{align}
where $\FT{f}^{\RM{F*}}$ is the complex conjugate of $\FT{f}^{\RM{F}}$
and $\delta$ the Dirac delta function.

\begin{figure}
\centering
  \includegraphics[width=0.85\linewidth]{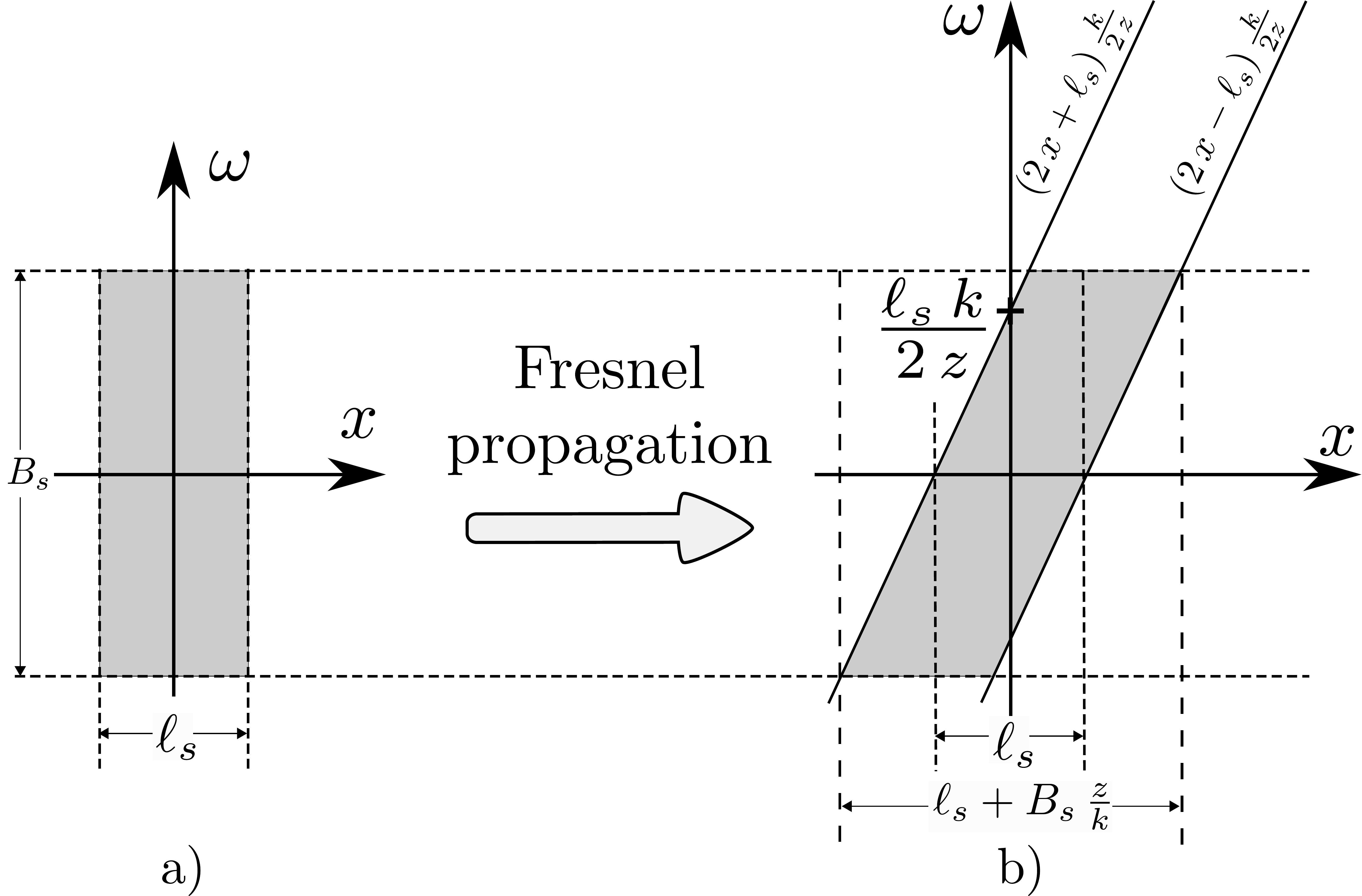}
  \caption{Transformation of the \sbp  in the Wigner domain after a Fresnel transform.}
  \label{fig:SWF}
\end{figure}

As a multiplication in Fourier domain is a convolution along the space
dimension in the Wigner domain, the Fresnel propagation corresponds to
a horizontal shearing of the signal \sbp \cite{Lohmann1993} as depicted in
\Fig{fig:SWF}.b. As stated in \cite{Stern2006}, to prevent loss of
information, the propagated signal must be sampled on an area of width $\ell_{s} + z \,\frac{B_{s}}{k}$   that encompasses all its support in the Wigner domain.

\begin{figure}
\centering
  \includegraphics[width=0.75\linewidth]{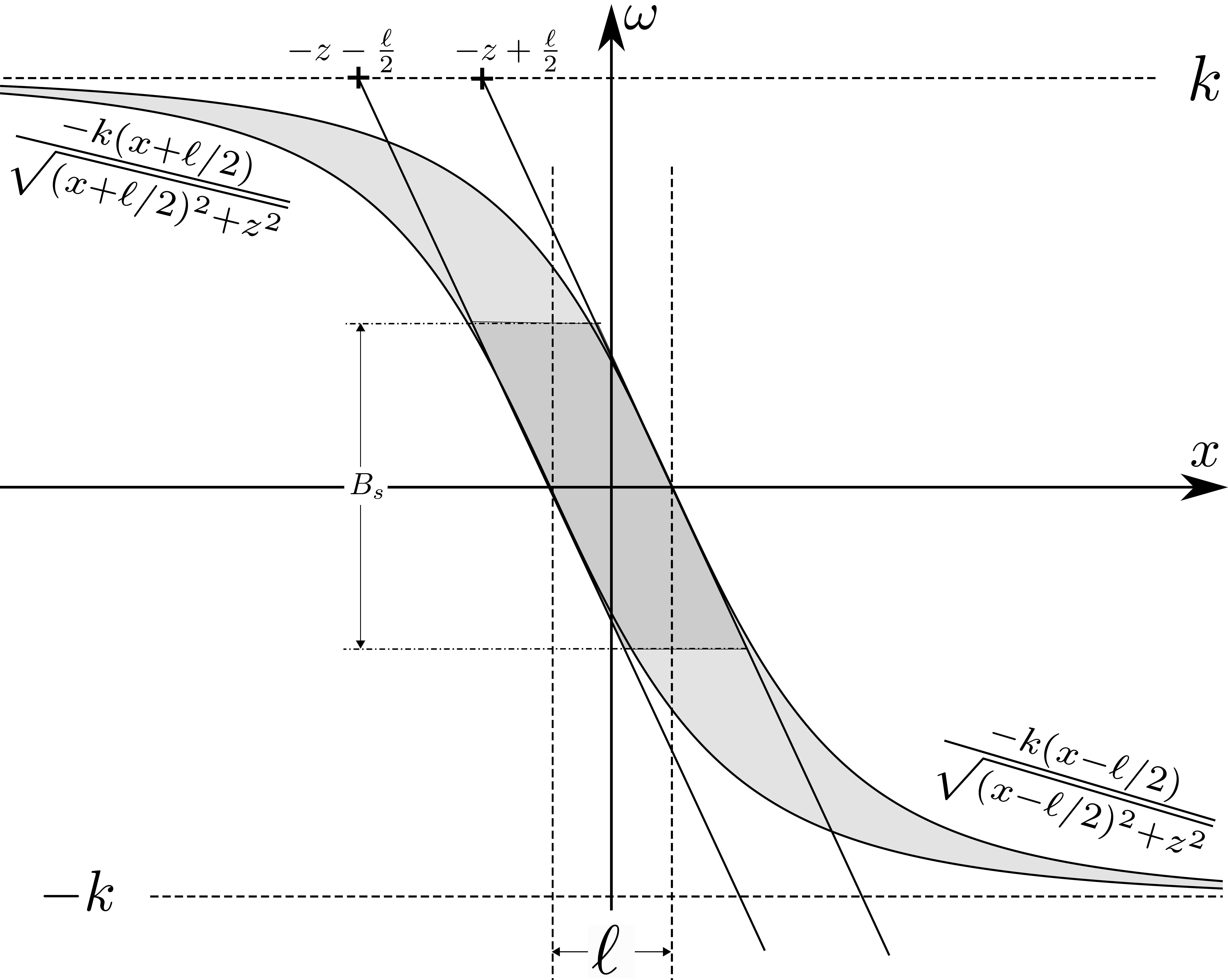}
  \caption{Space-bandwidth product (light gray) probed by a lensless
    setup of detector size $\ell$ under a normal illumination
    ($\theta=0$). The darker area represents the \sbp under paraxial
    approximation (when the bandwidth of the probed sample is much smaller than  $k$). }
  \label{fig:SWAS}
\end{figure}

Conversely, as in practice the propagated signal is sampled on a rectangular area
in the Wigner domain of width $\ell$, the lensless holographic setup probe a region of
the sampled sheared in the opposite direction as depicted on the dark
grey area of \Fig{fig:SWAS}. This means that under Fresnel approximation for a band-limited sample of angular bandwidth $B_s$, the 
\fov of a setup with a detector of width $\ell$ is  $\ell_1^\RM{F}\times \ell_2^\RM{F}$ with:
\begin{equation}
  \label{eq:FresnelFoV}
    \ell_i^\RM{F} =  \ell_i+ z\,\frac{B_s}{k}\,.
\end{equation}

As a  consequence of the shearing of the space-bandwidth product, the bandwidth and thus the resolution of a lensless setup varies within the \fov \cite{Fournier2010,Kelly2009,Hao2011}. 
The
resolution is coarser at the center (with a resolution of
$r_{\textrm{center}} = \frac{\lambda\,z}{2\,\ell}$ at best) than on the edges of the \fov (resolution of $r_{\textrm{edge}} = \frac{4\,\pi}{B_s}$ at best).

\subsection{Angular spectrum space-bandwidth product}
Wigner distribution function of the angular spectrum kernel \Eq{eq:FAS} is impossible to estimate analytically. However, its support in the Wigner domain can be estimated using the instantaneous frequency of the angular spectrum kernel expressed in space \Eq{eq:SASS}:
\begin{align}
  \xi_{i}(\V{x}) &= \frac{\D{\Arg(h^\RM{AS}(\V{x}))}}{\D{x_{i}}}\,,\\
                 & = k\left(\frac{x_{i}}{\sqrt{\Norm{\V{x}}^2+ z^2}} -\sin\left(\theta_{i}\right) \right)  \,,
\end{align}
where $\Arg(h)$ is the complex argument of ${h}$ and $\xi_{i}(\V{x})$
is the instantaneous frequency along the dimension $i$ and at the
position $\V{x}$. In 1D (taking $x = x_1$ and $x_2=0$):
\begin{equation}\label{eq:ASinstanfreq}
  \xi(x) =  k\left(\frac{x}{\sqrt{x^2+ z^2}} -\sin\left(\theta\right) \right)\,.
\end{equation}
The propagation kernel $h$ is a chirp and its instantaneous
frequency increases monotonically with $x$. This means that the probed 
bandwidth of a lensless setup varies within the
\fov. At a given position in the sample plane, the bandwidth probed by the setup is bounded by the instantaneous frequencies at each edges of the sensor. As a consequence, the \sbp of a lensless setup is the area between
$\xi(x+\frac{\ell}{2})$ and $\xi(x-\frac{\ell}{2})$ as
depicted \Fig{fig:SWAS}. It is:
\begin{align}
    \int_{-\infty}^{+\infty}  \xi\left(x+\frac{\ell}{2}\right)- \xi\left(x-\frac{\ell}{2}\right) \D{x} = 2\,k\,\ell\,,
\end{align}
that corresponds to a number of degrees of freedom of $2\,\frac{\ell}{\lambda}$: the highest
number  of degrees of freedom it is physically possible to record on a detector of size
$\ell$. The incidence angle shifts the \sbp along the angular frequency axis without
changing its shape. The center of the \fov{}  $\V{x}^c$ is given by:
\begin{equation}\label{eq:center}
  \xi(-x^{c}) =0 \iff  x^c = -z\, \tan(\theta)\,,
\end{equation}
as it is expected from geometrical optics.

For low spatial frequencies (\ie within the
paraxial approximation),  \sbp of AS and Fresnel (in dark gray on
\Fig{fig:SWAS}) are identical. 

To perfectly explain the propagated wave in the detector plane, one
has to model the sampled a rectangular area in the Wigner domain that encompasses all the \sbp of the angular spectrum. That is on a bandwidth of  $B = 2\,k$ centered on $k\,\sin(\theta) $. Along the
space dimension, this \sbp spans the whole axis implying an infinite \fov: $\ell^\RM{AS}= \infty$.  This
is a consequence of the infinite size in space of the AS kernel \Eq{eq:SASS}.
%
%
%
%

Assessing the size of the propagation kernel in space also amounts to
estimate the smallest sampling rate of the Fourier space needed to
perfectly sample $\FT{h}^\RM{AS}$.  It is estimated by computing its
highest instantaneous frequencies along both dimension:
\begin{equation}
  \tau_{i}(\V{\omega}) =
  \frac{\D{\Arg(\FT{h}(\V{\omega}))}}{\D{\omega_i}}\,.
\end{equation}
The instantaneous frequency $\tau_1$ (resp. $\tau_2$) is infinite for
$\omega_1=k \,\left( \sign(\theta_1)+ \sin(\theta_1)\right)$ (resp.
$\omega_2=k \,\left( \sign(\theta_2)+ \sin(\theta_2)\right)$).  The
diffraction patterns are thus infinitely extended in the detector
plane highlighting the bandlimited nature of $\FT{h}^\RM{AS}$.
However, $\FT{h}^\RM{AS}$ is still approximately well sampled as long
as the phase difference of $\FT{h}^\RM{AS}$ between two consecutive
frequels is lower than $\pi$ \cite{Voelz2009}. This is true when the \fov
$\ell' \times \ell'$ verifies :
\begin{align}
  & z \,\sqrt{n^2\lambda^{-2} - \left( n\,\lambda^{-1} - \ell'^{-1}\right)^2 }
    \leq 1\,,\\
  \label{eq:ltheo}
  & \ell' \geq \lambda \left(n-\sqrt{n^2-\left(\frac{\lambda }{z}\right)^2}\right)^{-1}\,.
\end{align}

The theoretical \fov given by  \Eq{eq:ltheo} is
too large, much larger than the effective \fov of the setup
in practice. As an example, $\ell'=3\,$m for the parameters used in
\cite{Rostykus2018}: $\lambda=681\,$nm and
$z=1.02\,$mm.  However several physical phenomena decrease the size of
diffraction pattern detectable in practice, narrowing the \fov.

\section{Narrowing the field of view}\label{sec:fov}
As the propagation is modeled as a convolution in space, every objects outside of the detector field but closer to its edge than the
half of the size of the propagation kernel in space will have an influence on
the propagated wave $w$ in the detector support. To take into an
account the fact that the effective propagation kernel may no be
symmetrical, we define its  size  along the dimension $i$ as the sum of its radius  to the left $p_{i}^{-}$ and to the right $p_{i}^{+}$.
For each phenomenon, namely: the spatial and the spectral coherence
and the detector noise and quantification, we define the half-width of the
diffraction pattern $p^\RM{spa}$, $p^\RM{spe}$, and $p^\RM{noise}$
respectively.  Along dimension $i$, the overall half-width of a propagation kernel is given by
\begin{align}
    p_i^- & = \min(p_i^\RM{-spa}\,,p_i^\RM{-spe}\,,p_i^\RM{-noise}) \,,\\
    p_i^+ & = \min(p_i^\RM{+spa}\,,p_i^\RM{+spe}\,, p_i^\RM{+noise})\,.
\end{align}
The final \fov of the setup is then $\ell'_1\times\ell'_2$ with:
\begin{equation}\label{eq:finalfov}
    \ell_i' = \ell_i +  p_i^- + p_i^+\,.
\end{equation}

\subsection{Measurements noise}
\label{sec:noise}

\begin{figure}
\centering
  \includegraphics[width=0.5\linewidth]{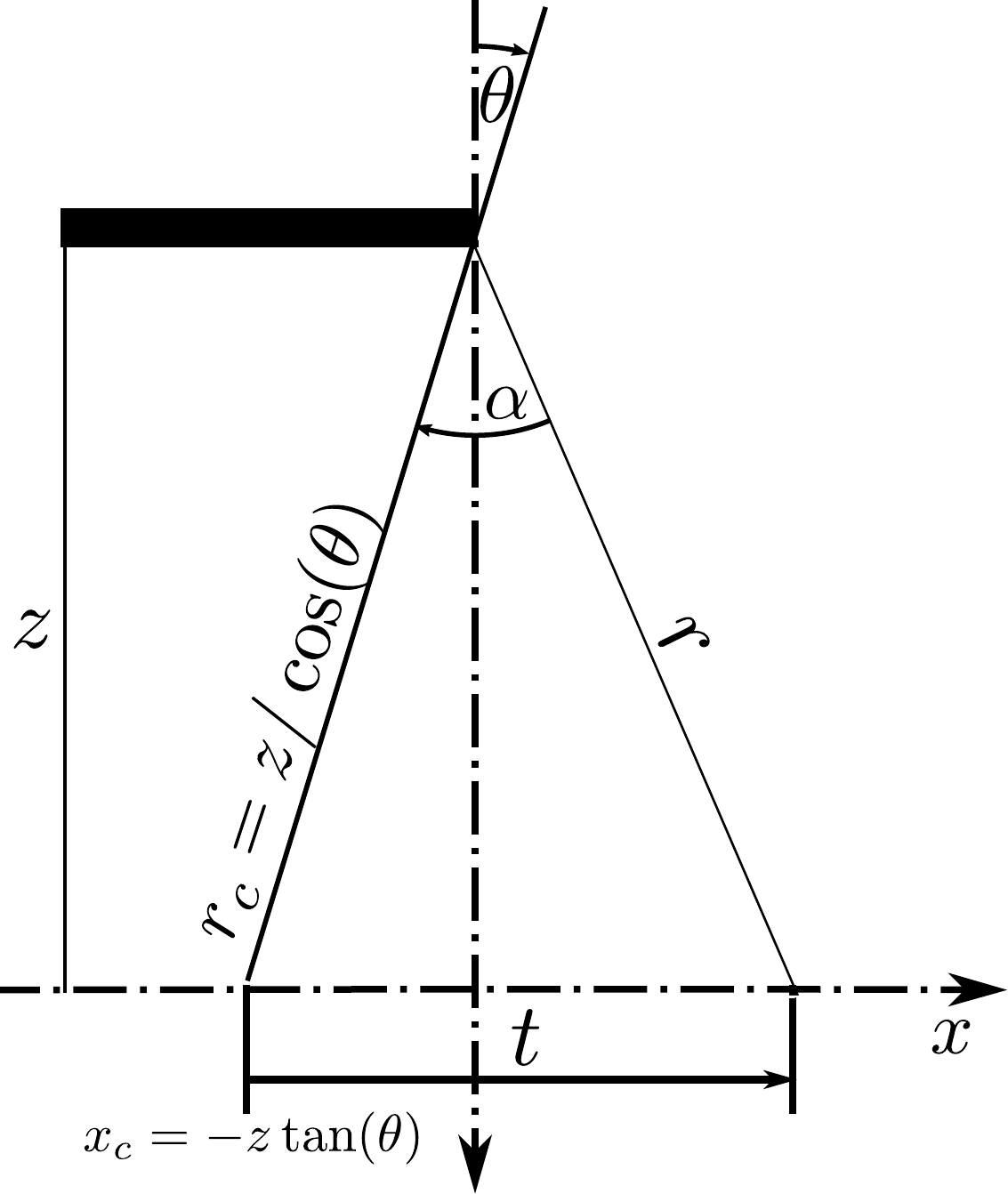}
  \caption{Scheme of the diffraction by a half plane under illumination with an incidence of $\theta$ as modeled in \Sec{sec:noise}. The geometrical shadow boundary is at  $x_{c}$. 
  }
  \label{fig:Edge}
\end{figure}

The extent of the diffraction kernel can  be
defined as the area where its intensity is over the detector detection
limits. This can be characterized as the distance $p$ from the center of a diffraction kernel after which  the fringes are no longer detectable by the sensor.

\begin{figure}
\centering
  \includegraphics[width=0.6\linewidth]{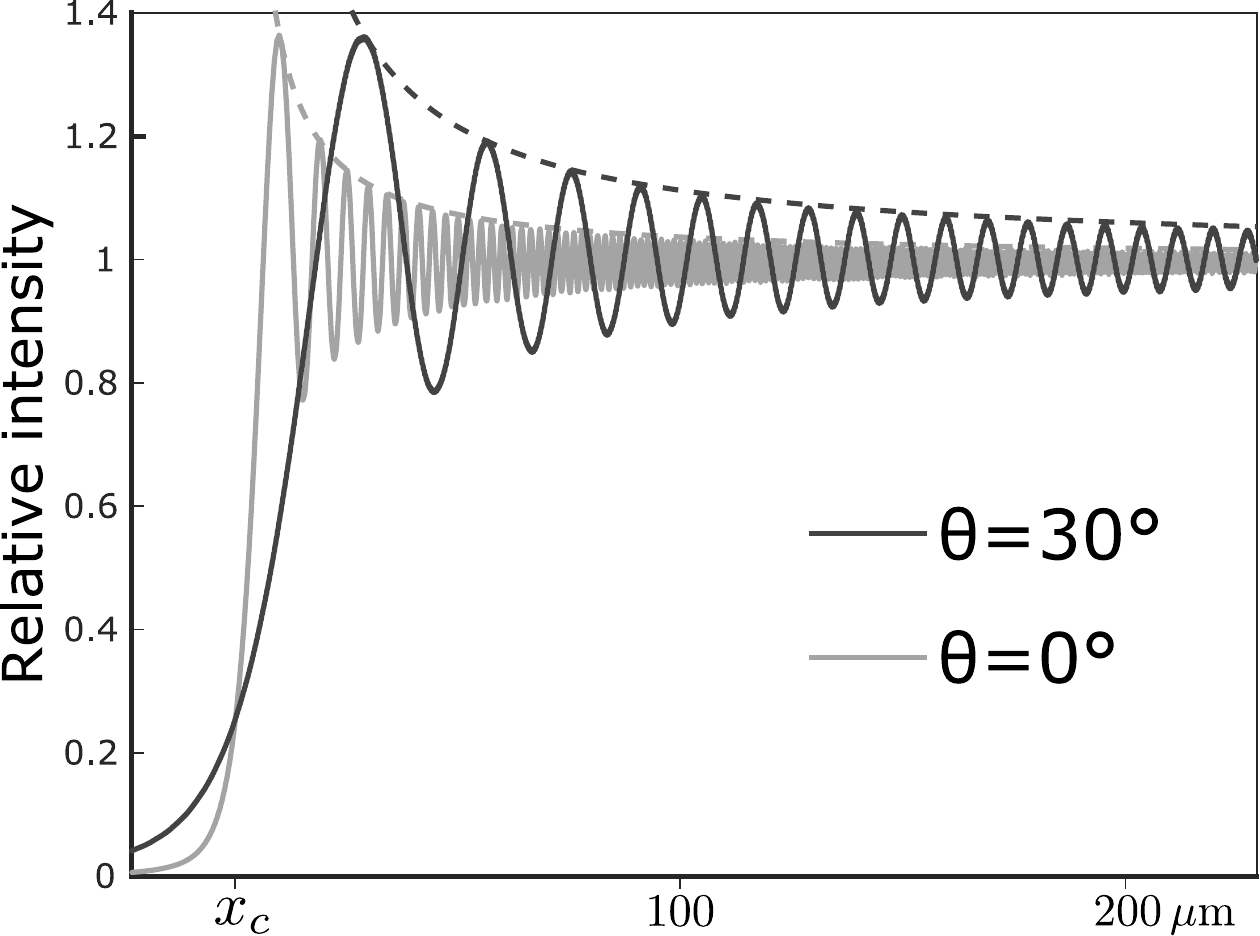}
  \caption{Intensity as a distance from geometrical shadow boundary   $x_{c}$ of the diffraction of a half plane at a
    distance of $z=250\,\micron$ for an
    illumination of wavelength $\lambda=530\,\nm$ and an incidence $\theta_{1} = 0^{\circ}$  (grey solid line) and $\theta_{1} = 30^{\circ}$ (black solid line) both with their upper bound (dashed lines) computed according to \Eq{eq:UpperBound}. }
  \label{fig:EdgePattern}
\end{figure}

In the detector plane, the contrast of a diffraction pattern
decreases with the distance from its center. As it scatters more
power, the larger the object is, the wider  its diffraction
pattern is. The largest diffraction pattern is given by
a half plane:
\begin{equation}
  o(x_1,x_2)= \left\{
    \begin{array}{ll}
      0 &   \text{if }
          x_1 \le 0,\\
      1 & \text{otherwise} \,.
    \end{array}
  \right.
\end{equation} 
Along the $x_{1}$ axis, the diffraction by a half plane under inclined illumination is given by \cite{Sheppard1991}:
\begin{align}
  w(r,\alpha) =  &\exp\left(-\jmath\pi/4\right)\sqrt{\frac{ 1 + \cos(\alpha) }{\pi\,\cos(\theta)\,\left( \cos(\theta+ \alpha) + \cos(\theta)\right)}} \nonumber \\
  & \times \exp\left(\jmath \,k\,r\,\cos(\alpha)\right)\,F\left( \sqrt{2\,k\,r}\, \sin\left(\frac{\alpha}{2}\right)\right)\,,
\end{align}
where $(r,\alpha)$ are polar coordinates as depicted on \Fig{fig:Edge}. $F$ is the Fresnel integral under the form:
\begin{equation}
  \label{eq:FresnelInt}
  F(x) = \int_{x}^{\infty} \exp\left(\jmath\,t^{2}\right) \,\D{t}\,.
\end{equation}
The geometrical shadow of the edge is $x_{c}= z\,\tan(\theta)$ and its
distance from the edge is $r_{c}  = z / \cos(\theta)$. 
In an in-line setup, the intensity for $t>0$ ($\alpha >0$) is
approximately bounded  by (see \Fig{fig:EdgePattern} for two examples) :
\begin{align}\label{eq:UpperBound}
  \Abs{w(x_c + t,x_{2})}^2 \lesssim 1+  \frac{\sqrt{2\,\left(r_{c} + t\,\sin(\abs{\theta})\right)}}{\sqrt{\pi\,k}\,t\,\cos(\theta)}\,. 
\end{align}

 The half-width of the diffraction pattern to the left $p^-$ is  the 
 distance $t$ from $x_{c}$ after which the fringes are no longer
 detectable. 
The threshold $\eta$ below  which the fringes are no longer visible
is defined as the smallest effective quantification level relatively
to illumination mean intensity $I_0$.
\begin{equation}
  \eta\approx (\RM{SNR})^{-1/2} 
  =\max\left(\frac{\sigma}{I_0}, \frac{\zeta}{I_0}\right)\,,
\end{equation}
where $\sigma$ is the mean standard deviation of the noise and $\zeta$
is the quantification level of the camera.
The half-width of the diffraction pattern to the right $p^+$ is given by the symmetry with respect to the sensor (that is changing the sign of $\theta$).  As the bound in \Eq{eq:UpperBound} depends only on $\abs{\theta}$, we have $p_i^{+\RM{noise}}=p_i^{-\RM{noise}} = p_i^{\RM{noise}}$.  Given the threshold $\eta$, the half-width of the diffraction pattern along dimension $i$ is $ p_i^{\RM{noise}}$, such that:
\begin{align}
& \frac{\sqrt{2\,\left(r_c  + p_i^{\RM{noise}}\,\sin(\abs{\theta_i})\right)}}{\sqrt{\pi\,k}\,p^{\RM{noise}}_i\,\cos(\theta_i)}  = \eta  \\
 & p_i^{\RM{noise}} = \frac{\sqrt{ 2\,z\,
      k\,\pi\,\eta^{2}\,\cos(\theta_i) + \sin^{2}(\theta_i)} + \sin(\abs{\theta_i})}{ k\,\pi\,\eta^{2}\,\cos^{2}(\theta_i) }\,.
\label{eq:NoiseBnd}
\end{align}

As an example,  for the parameters used in \cite{Rostykus2018} ($\lambda=681\,$nm and
$z=1.02\,$mm, $\theta_1=9^\circ$) and a camera  of width $\ell_1
=5.3 
\,\textrm{mm}$ and $20\,$dB of noise ($\eta=10^{-2}$), the estimated bound is $p_i^{\RM{noise}} = 911$\micron{} leading to a width of the field of view of $\ell_1' = 7.1\,\textrm{mm}$ much more tractable than $\ell'
=3\, \textrm{m}$ given by \Eq{eq:ltheo}.

Let us notice that in detection applications such as \cite{Soulez2007a} scatterers can be detected in an even larger \fov as the noise is averaged on a large number of pixels increasing the effective $\RM{SNR}$.

\subsection{Partial coherence}

\begin{figure}\centering
  \includegraphics[width=0.7\linewidth]{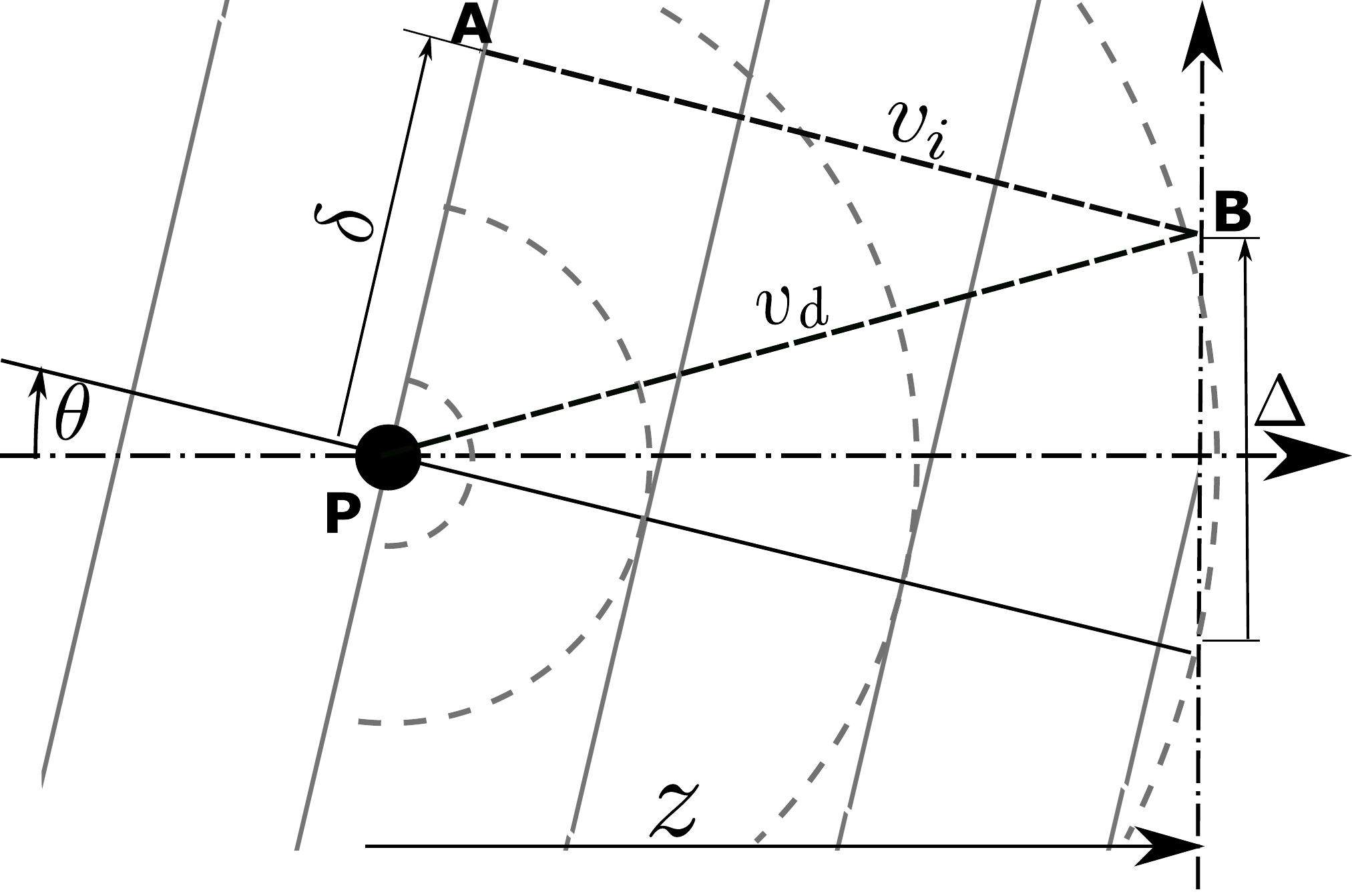}
  \caption{The extend of a diffraction pattern boundary is the point where optical path difference between the diffracted wave $v_{d}$ and the illumination wave $v_{i}$ is larger than the coherence length.}
  \label{fig:SpatialCoherence}
\end{figure}

In the previous sections, we always consider illumination light as a single monochromatic plane wave.  
However, in practice any real source must have a finite size  and a finite \changes{spectral bandwidth. This partial coherence affects the size of the diffraction patterns as it decreases the visibility of the fringes and blurs out intensity high spatial frequencies. As the spatial frequencies of the fringes increase with the distance from the center of the diffraction pattern, this  effectively reduces its extent.} It is possible to assess the extension of the diffraction pattern by considering the Born approximation model of inline holography where the diffraction pattern is generated by the interferences between the illumination wave $v_i$ and the  wave $v_d$ diffracted by a scatterer P as depicted  on \Fig{fig:SpatialCoherence}. \changes{On   the detector plane, at a distance  $\V{\Delta} = [\Delta_1,\Delta_2]\Tt$  from the center  of the diffraction pattern, the  coherence of these two waves can be described by their mutual coherence factor}\,\cite{GoodmanStatisticalOptics}\,:
\begin{align}
\mu(\V{\Delta})    = \frac{\Avg{ v_i(\V{\Delta},t)\,v^*_d(\V{\Delta},t)}_t}
                        { \sqrt{\Avg{v_i(\V{\Delta},t)\,v^*_i(\V{\Delta},t) }_t \,\Avg{v_d(\V{\Delta},t)\,v^*_d(\V{\Delta},t) }_t }}\,
\end{align}
\changes{where  $\Avg{v(t)}_t$ is the average over time. When $\V{\Delta}=0$ (at the center of the diffraction pattern) illumination wave and diffracted wave coincide and $\Abs{\mu(\V{\Delta})} $ is maximum.  When $\Abs{\mu(\V{\Delta})}=0$ the two waves are mutually incoherent and no longer  interfere. }

\changes{To simplify further computations, we set an  upper bound estimate on $\mu(\V{\Delta})$ that is separable along both lateral axis giving the maximal possible extension of the diffraction pattern:}
\begin{align}
\Abs{\mu(\V{\Delta}) }\leq \Abs{\mu_1(\Delta_1)} \,  \Abs{\mu_2(\Delta_2)} \,.
\end{align}

\changes{Along dimension $i$, the coherence factor $\mu_i(\Delta_i)$ can be inferred from both spectral and spatial coherence of the illumination wave (supposed quasi-homogeneous\,\cite{GoodmanStatisticalOptics})  given by its complex degree of coherence\,:}
\begin{align}
\gamma(\V{\delta},\tau)    = \frac{\Avg{ v_i(\V{x}, t)\, v^*_i(\V{x}+\V{\delta}, t+\tau)}_{\V{x},t}}
                                    { \Avg{v_i(\V{x},t)\,v^*_i(\V{x},t) }_{\V{x},t}}\,.
\end{align}    
\changes{Using the notation given on   \Fig{fig:SpatialCoherence}, the distance between P and A is $ \delta_i^{AP}=\frac{\Delta_i}{\cos{(\theta_i)}}$ the relative time delay between B and A (for $v_i$), and P and B (for $v_d$) are:}
\begin{align}
    \tau^{AB}_i &=   \frac{n\,z}{c\,\cos(\theta_i)}\,,\\
    \tau^{PB}_i &=  \frac{n}{c}\left(\sqrt{z^{2}  + \left(z\,\tan(\theta_i) + \Delta_i \right)^2} + \Delta_i \,\sin(\theta_i)\right)\,.
\end{align}
\changes{This gives the  mutual coherence factor between $v_i$ and $v_d$ at position $\Delta_i$:}
\begin{align}
    \mu_i(\Delta_i) &= \gamma( \delta_i^{AP},\tau^{PB}_i - \tau^{AB}_i)\,,\\
    & = \gamma\left(\frac{\Delta_i}{\cos{(\theta_i)}} , \frac{n}{c}\left( \sqrt{z^{2}  + \left(z\,\tan(\theta_i) + \Delta_i \right)^2}-
  \frac{z}{\cos(\theta_i)} + \Delta_i \,\sin(\theta_i)\right)\right)\,.
\end{align}
\changes{This relation will be used to determine the half-width of the diffraction pattern in the two limiting cases: (i) a spatially coherent illumination ($\gamma({\delta},\tau) = \gamma(0,\tau),\ \forall {\delta}$ )
and (ii) a spatially incoherent quasi-monochromatic illumination ($\gamma({\delta},\tau) = \gamma({\delta},0),\ \forall {\tau}$ ).}



\subsubsection{Spectral coherence}

In the case where the source is spatially coherent, the self coherence of the illumination wave is described by its coherence
length $L_c$ that  depends on the shape and the spectrum of the illumination light.
We estimate the half-width of the diffraction
pattern as the distance  where the  optical path difference $c\, (\tau^{PB}_i- \tau^{AB}_i)$ is equal to the
coherence length $L_c$. 
The diffraction pattern is not symmetric and along dimension $i$ its half-width to the left $p_i^{+\RM{spe}}$ and to the right $p^{-\RM{spe}}$ are:
\begin{align}
  p_i^{+\RM{spe}} &= \frac{L_c}{n\,\cos^2(\theta_i)}\left(\sqrt{1 + \frac{2\,n\,\cos(\theta_i)\,z}{L_c}}
      + \sin(\theta_i)\right)\,.\label{eq:SpeCohe}\\
  p_i^{-\RM{spe}} &= \frac{L_c}{n\,\cos^2(\theta_i)}\left(\sqrt{1 + \frac{2\,n\,\cos(\theta_i)\,z}{L_c}}
      - \sin(\theta_i)\right)\,.
\end{align}

Again,  for the parameters used in \cite{Rostykus2018} ($\lambda=681\,$nm and $z=1.02\,$mm, $\theta_1=9^\circ$ and  camera  width $\ell_1=5.3\,\textrm{mm}$), 
and a source coherence length of $L_c = 1\,$mm, the estimated bounds are $ p_i^{+\RM{spe}} = 1.9\,$mm and $p_i^{-\RM{spe}} = 1.6\,$mm  leading to a width of the field of view of $\ell_1' = 8.8\,\textrm{mm}$.

\subsubsection{Spatial coherence}
In the case of a quasi-monochromatic spatially incoherent  source, the visibility of the fringes and hence the width of the diffraction pattern depends on the source coherence factor in the illumination plane $\mu_s(\V{\Delta}) = \gamma(\V{\Delta},0)$. Following the Van Cittert-Zernike theorem, this  coherence factor is the Fourier transform of the source brightness distribution. Depending on the shape of the coherence factor,  one can define  the half-width $p_i$ as the  half-width at half-maximum (HWHM) along dimension $i$. When  the source brightness distribution has a circular symmetry, the half-width $p_i$ can be also defined as the radius of the coherence area.

For a uniformly bright circular source of angular radius $\alpha$, the coherence factor of the illumination wave is a cardinal Bessel function of order 1 \cite{GoodmanStatisticalOptics}.  Its HWHW is:
\begin{equation}\label{eq:SpaCohe}
     p_i =   0.35 \frac{\lambda}{n\,\tan(\alpha)\cos(\theta_i)}
\end{equation}
The coherence area of this circular source is $A_c = \frac{\lambda^{2}}{\pi \tan^2(\alpha)}$. Therefore, for an incidence $\V{\theta}$, the radius of source coherence area is:  
\begin{equation}
     p_i =  \frac{1}{\pi} \frac{\lambda}{n\,\tan(\alpha)\cos(\theta_i)}\,,
\end{equation}
that is similar to the HWHM criterion ($1/\pi \approx 0.32$).



\section{Narrowing the bandwidth}\label{sec:bandwidth}

Even with a narrower  \fov $\ell_1'\times\ell_2'$ given by \Eq{eq:finalfov}, the number of degrees of freedom to perfectly represent the object can still be huge ($N = 2\,\ell'_1/ \lambda \times  2\,\ell'_2/ \lambda $). However it is possible to decrease this number of freedom  by estimating a narrower  effective angular bandwidth of the setup $B'< 2\,k$.

\begin{figure}
\centering
  \includegraphics[width=0.75\linewidth]{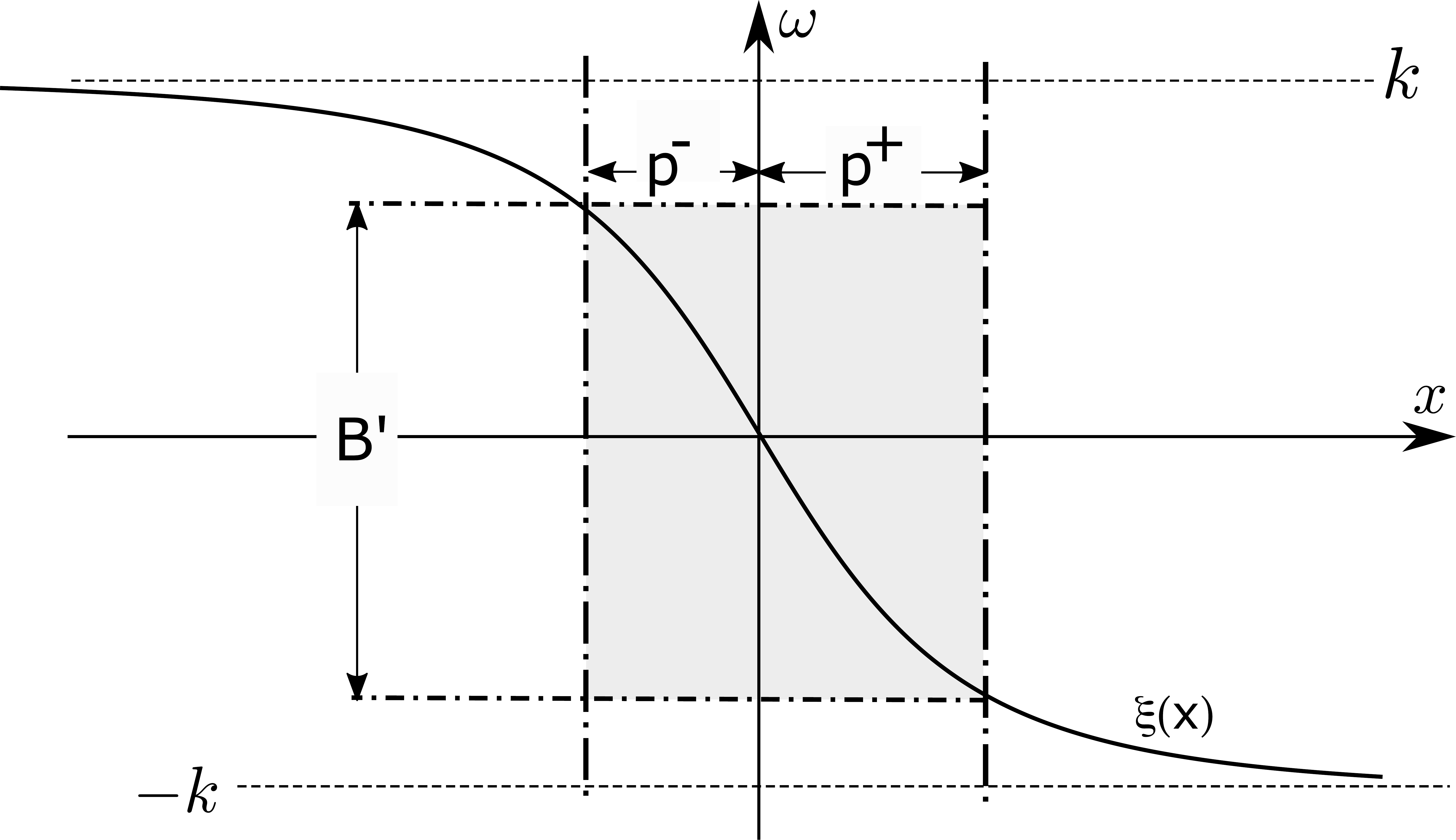}
  \caption{Effective bandwidth $B'$ estimated from the extent of the a diffraction pattern $p^- + p^+$.}
  \label{fig:SWASBandwidth}
\end{figure}
 
The effective angular bandwidth can be estimated as the highest angular
frequency of the  propagation kernel. As it is a frequency chirp, the highest
angular frequency of the propagation kernel is given by its instantaneous frequency at the farthest point from its center (that is at $p^-$ or $p^+$) as illustrated \Fig{fig:SWASBandwidth}:
\begin{align}
   B'_i &=   2\,\max\Big(\xi(p^-_i), \xi(p^+_i)\Big)\,,
\end{align}
with the instantaneous angular frequency $\xi(x)$ given in  \Eq{eq:ASinstanfreq}. 
\changes{This gives the spatial resolution $R_i$ along dimension $i$:
\begin{align}\label{eq:pixelpitch}
    R_i = \frac{4\, \pi}{B'_i} =      \frac{\lambda}{n} \left(\frac{p}{\sqrt{p_i^2+z^2}} +\sin(\abs{\theta_i})\right)^{-1} \,.
\end{align}
All waves have thus to be sampled along dimension $i$ with a maximum pixel pitch  of 
$\Delta_i = {R_i}/{2}$.}

In the example given \Sec{sec:noise} with $p_i^{\RM{noise}} = 911\,$\micron{}, this leads to a maximum pixel pitch of $\Delta_i =  511\,$nm instead of $\lambda/2 =  340\,$nm dividing the number of freedom by $2.25$.

In addition, setting $B' = \max(B_1', B_2')$ and $p = \max(p^-_1,p^+_1,p_2^-,p_2^+)$, the effective bandwidth $B'$ can  be used to determine whether or not the conditions of Fresnel approximation are fulfilled, namely if:
\begin{align}
    \left( \frac{B'}{2}\right)^2 \ll k^2 \Longleftrightarrow \frac{p^2}{p^2+z^2} + \sin(\abs{\theta})\ll 1\,.
\end{align}

In addition, it is also possible to derive from this bandwidth the effective numerical aperture of the setup. As the bandwidth along  each axis $B'_1$ and $B'_2$ can be different, the resolution of the setup is anisotropic.  Furthermore as explained in paragraph 3.B, the bandwidth of a lensless setup varies with the position within the \fov. As a consequence, we define $\NA_1(\V{x})$ and $\NA_2(\V{x})$ the numerical apertures along both axis at the position $\V{x}$ in the \fov. For the sake of simplicity, we consider only the case of normal incidence ($\theta_1=\theta_2=0$) and a square detector of size $\ell\times\ell$. In that case, the half width of the diffraction pattern is $p^-_1=p^+_1=p_2^-=p_2^+=p$.
When this size $p$ is smaller than the half-width of the detector $p<\ell/2$, the  numerical aperture is approximately constant across the \fov{} and $\NA_1 =\NA_2= n\,\frac{p}{\sqrt{p^2+z^2}}$. When $p>\ell/2$, numerical apertures along both axis vary across the \fov{}. They are minimal at the center of the \fov $\V{x}^c$ (as defined in \Eq{eq:center}) and maximal at its edges: 
\begin{align}
    \NA_i(\V{x})= n\,\frac{\min(\abs{x_i - x^c_i}+\ell/2, p)}{\sqrt{\min(\Norm{\V{x} - \V{x}^c + \ell/2}^2,p^2)+z^2}}
\end{align}

In a 3D imaging context, it is possible to derive from these numerical aperture, the \changes{two points resolution along the depth dimension}  of an in-line holography setup. In Born approximation, it is the minimal distance $d_z$ along the depth axis where two diffraction patterns can be disentangled. As the resolution, it varies within the \fov{}:
\begin{align}
    d_z(\V{x}) = \frac{n\,\lambda}{\max(\NA^2_1(\V{x}), \NA^2_2(\V{x}))}\,.
\end{align}


\section{Numerical experiments}

\subsection{Methodology}

\begin{figure}
\centering
  \includegraphics[width=0.75\linewidth]{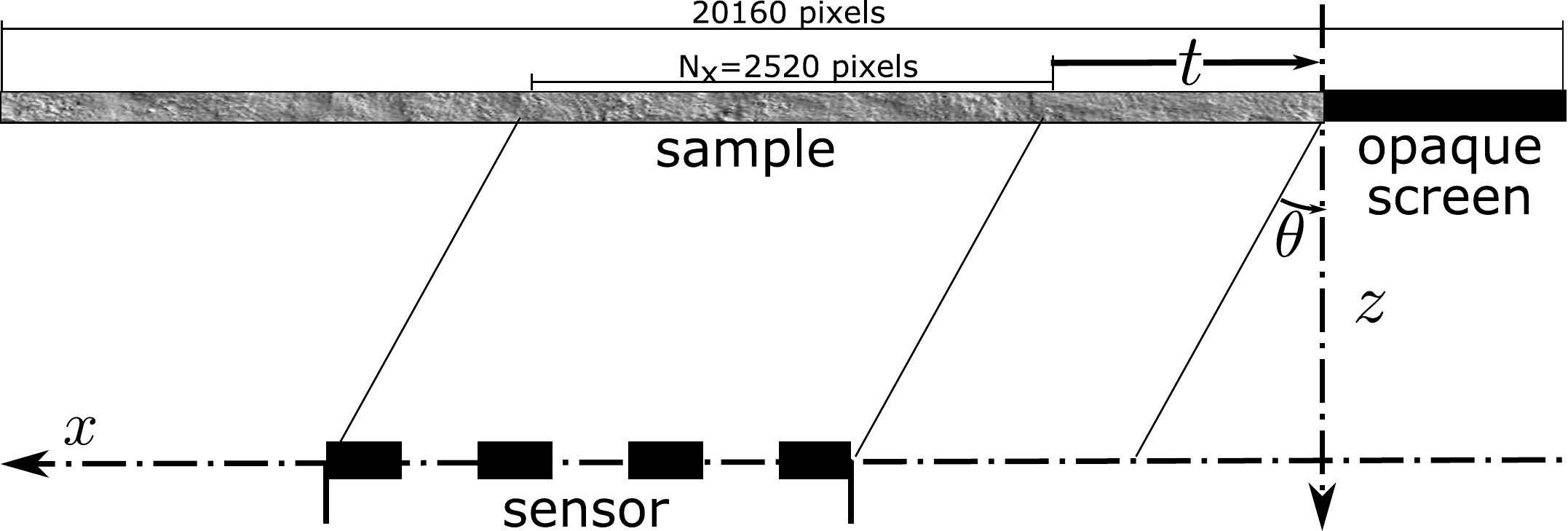}
  \caption{Scheme of the simulated setup where a part of the sample is masked by an opaque screen placed at a distance $t$ of the sensor edge projected in the sample plane along the direction of incident light.}
  \label{fig:Simu}
\end{figure}
 
\begin{figure*} 
    \centering
    \setlength{\tabcolsep}{0.005\linewidth}
    \begin{tabular}{cccccc}
       \includegraphics[height=0.18\linewidth]{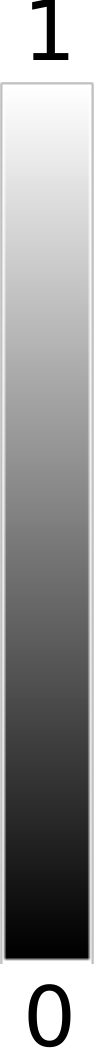}& \includegraphics[width=0.18\linewidth]{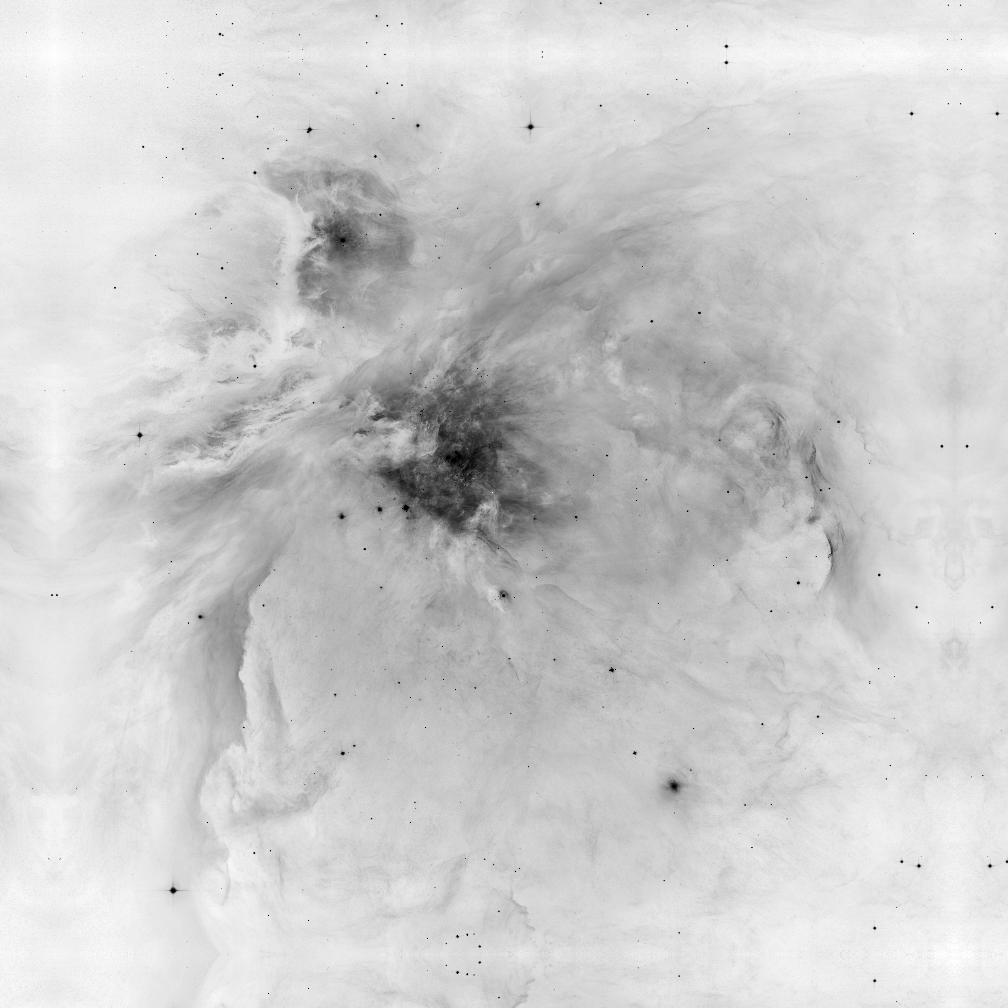} &
        \includegraphics[width=0.18\linewidth]{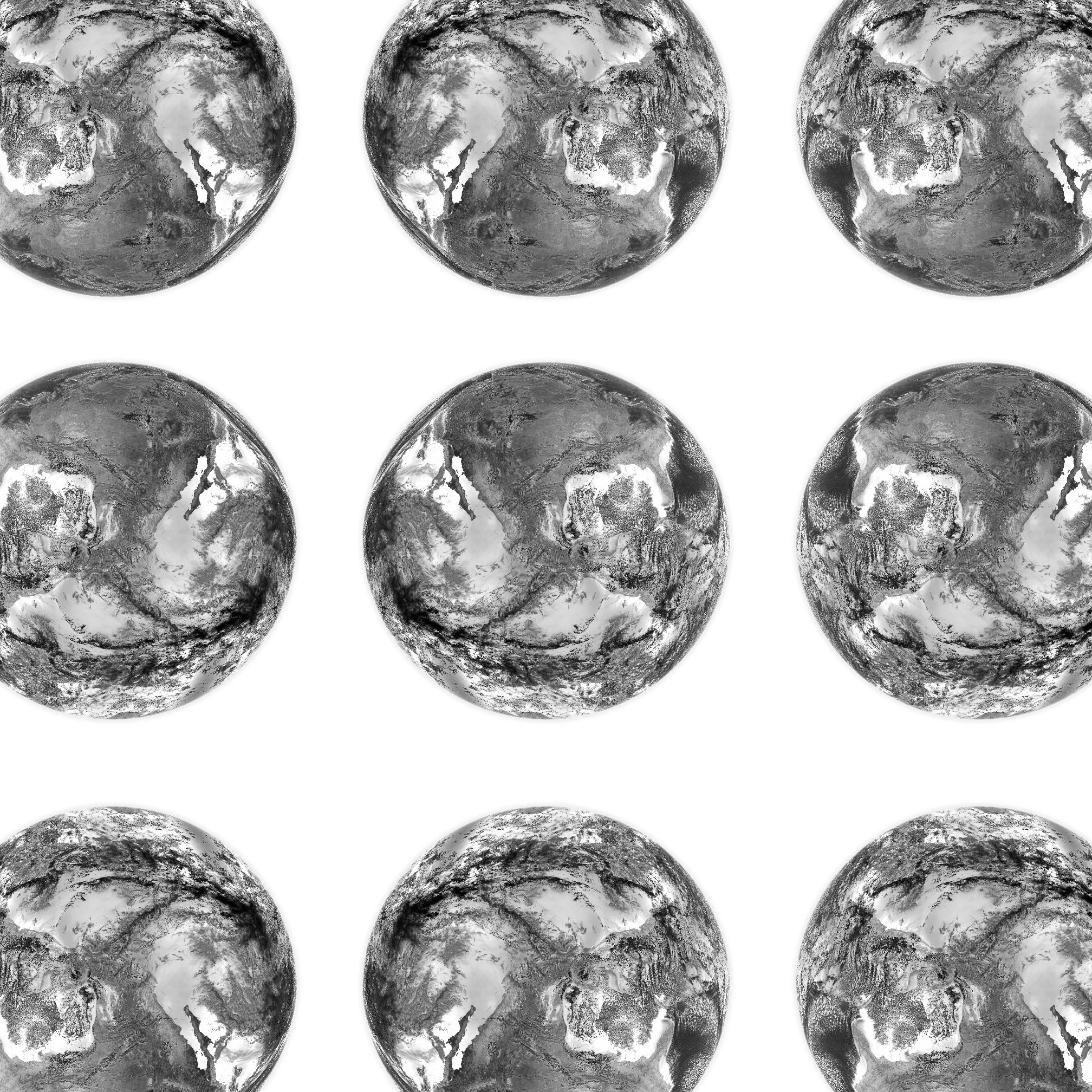} & \includegraphics[width=0.18\linewidth]{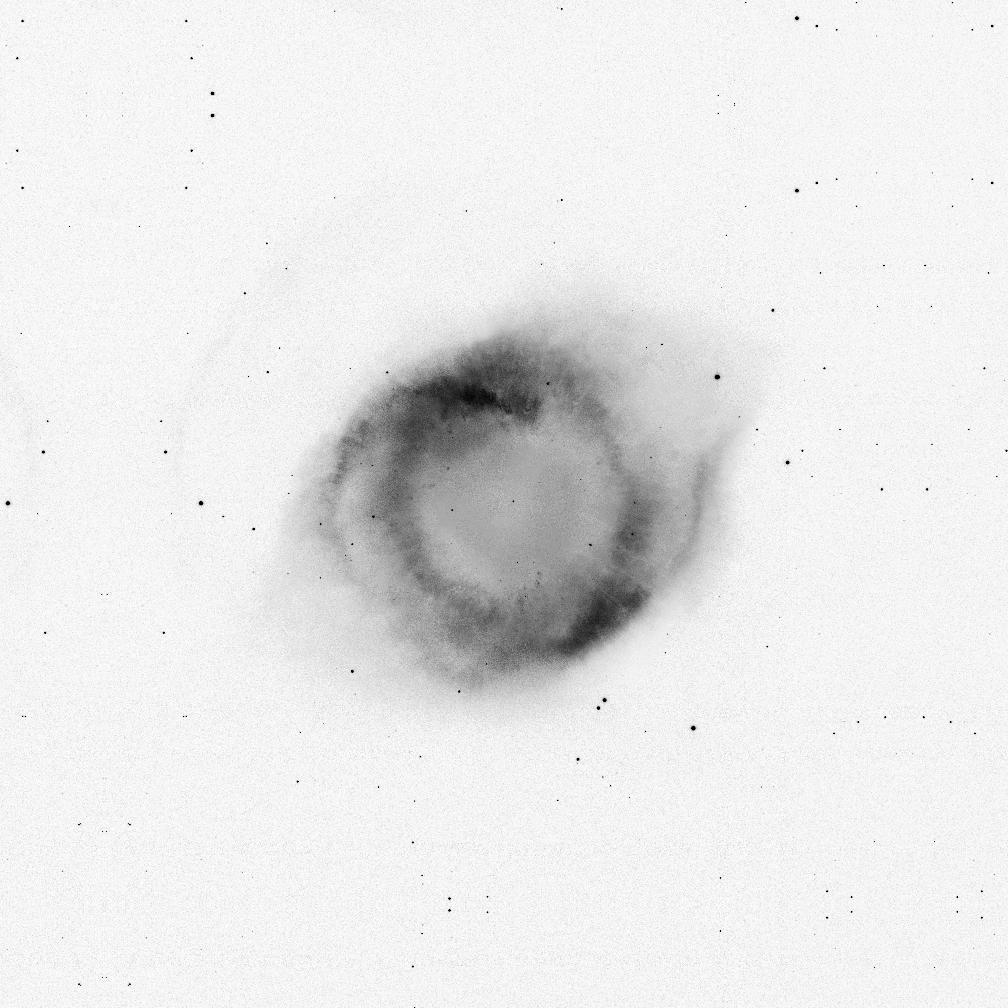} & 
        & \includegraphics[width=0.18\linewidth]{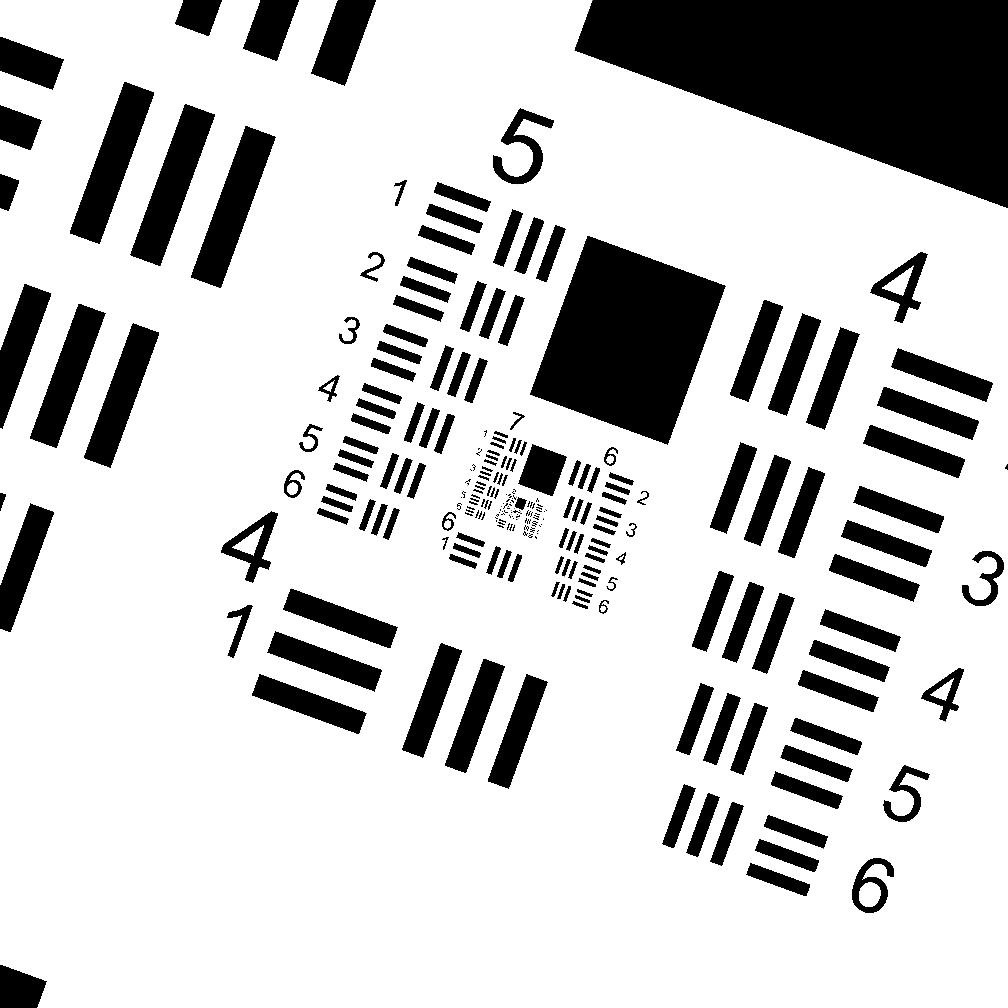} \\
       \includegraphics[height=0.18\linewidth]{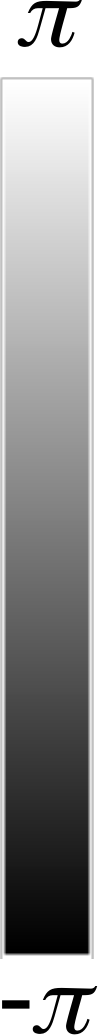}&
        \includegraphics[width=0.18\linewidth]{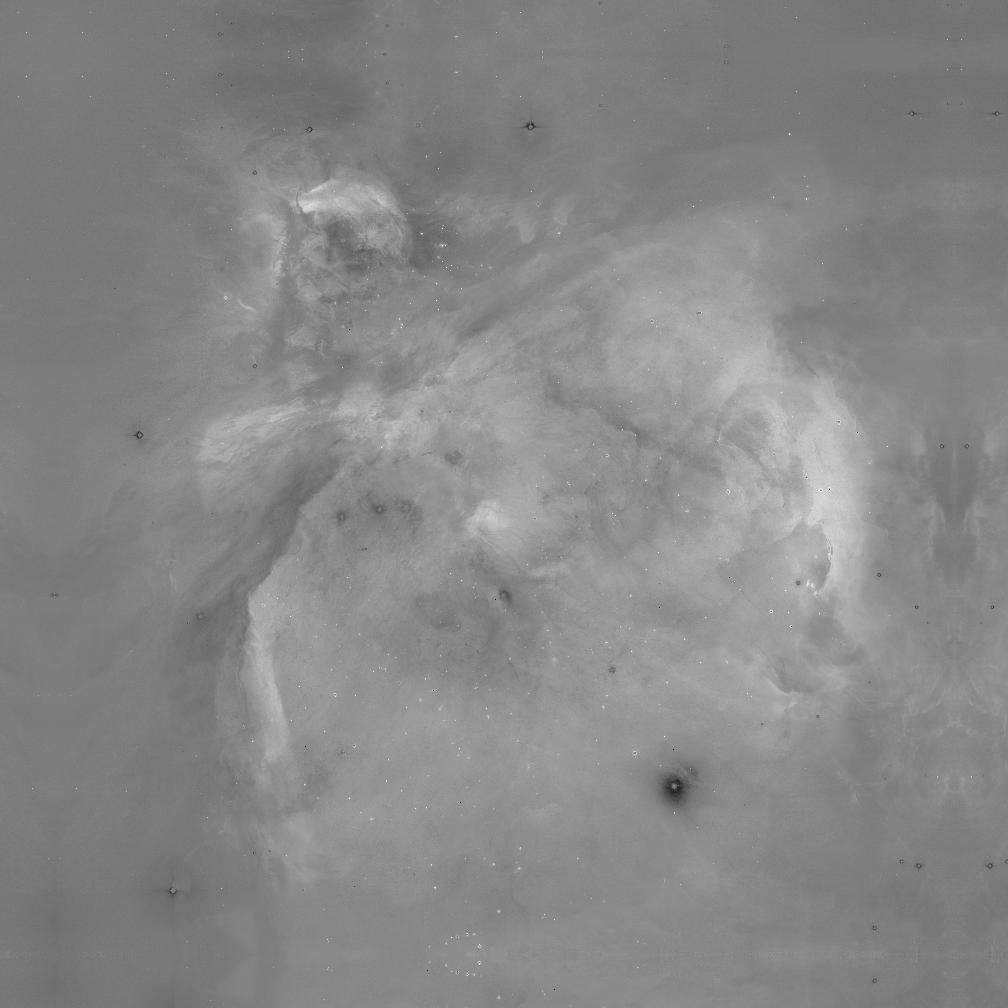} &
        \includegraphics[width=0.18\linewidth]{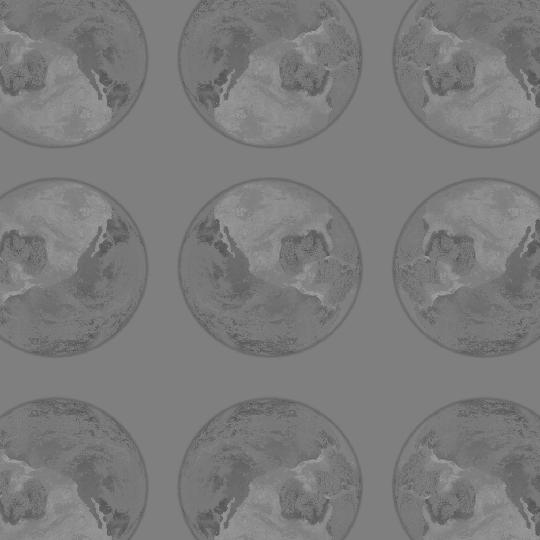} & \includegraphics[width=0.18\linewidth]{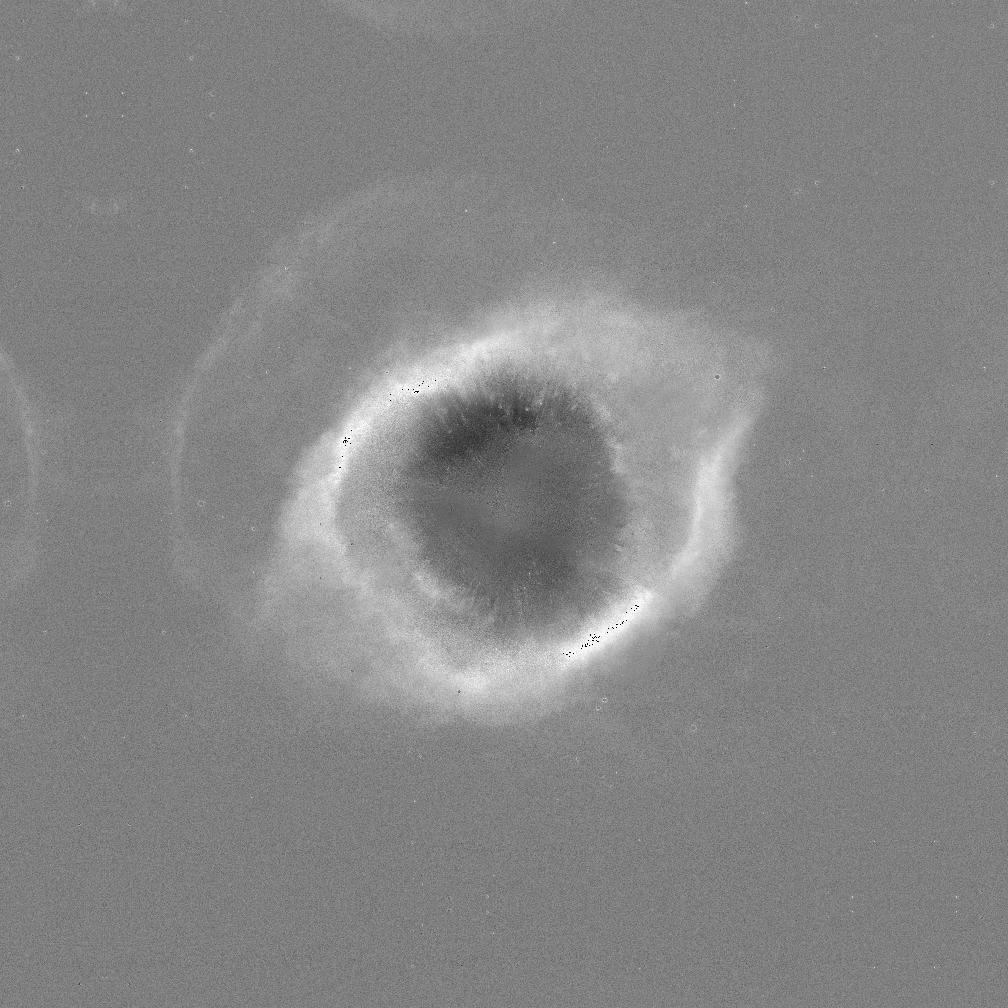} & \includegraphics[width=0.18\linewidth]{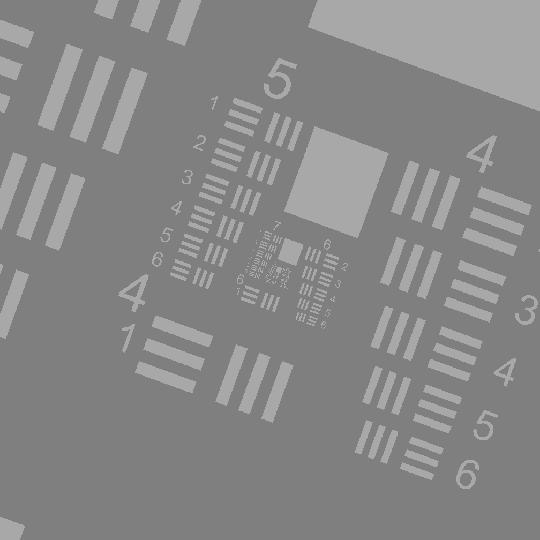} & \includegraphics[width=0.18\linewidth]{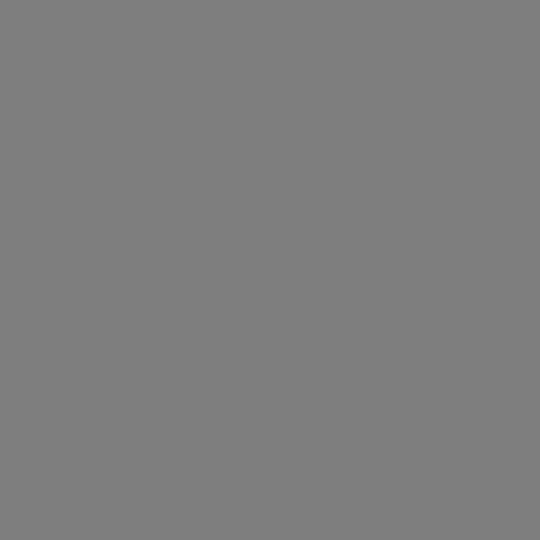} \\
       & (a) & (b) & (c) & (d) & (e) 
    \end{tabular}
        \caption{\label{fig:imageset} Transmittance (first row) and phase (second row) of the five $20160\times20160$ pixels images used to generate test dataset. (d) is a phase only USAF-1951 image and (e) is a transmittance only binary USAF-1951 image.}
\end{figure*}

We now illustrate the  usefulness of the bounds derived in the previous sections on simulations. We define a reference intensity $\V{r}$ with suitable noise and   coherence  properties. This reference is established by simulating the propagation over a large \fov \changes{of  $20160\times20160$ pixels while computing the modeling error $E$ in each experiments on a much smaller area ($2520 \times 2520$ pixels).}  
The effect of partial coherence is simulated using a Monte-Carlo method.  

In all experiments, we plot  the  modeling error   $E(t)$ (in dB) between this reference $\V{r}$  and the intensity $\V{s}$ computed when part of the sample is zeroed by an opaque screen as depicted on \Fig{fig:Simu}. The intensity $\V{s}(t)$ is simulated by propagating the masked sample exactly as the reference $\V{r}$  but without noise and coherence effects.
This error $E$  is computed as a function of the  lateral distance $t$ of the screen to the edge of the sensor projected at the sample depth parallel to the illumination incidence. It  is normalized by the error for the intensity  computed without the screen  $\V{s}_\infty$ (or when $t$ is very large):
\begin{equation}\label{eq:errors}
    E(t) = 10\,\log_{10} \Norm{\V{r} - \V{s}(t)}^2_2 -  10\,\log_{10} \Norm{\V{r} - \V{s}_\infty}^2_2 \,. 
\end{equation}
This error is maximum when $t=0$ showing the modeling error made when the  \fov it restricted to the size of the sensor and goes to $0$ as $t$ increase.


For each experimental condition the measured modeling error $E(t)$ is averaged over many experiments  with different objects and noise realizations. We use 5 different images that were randomly shifted to generate dozens of images used  in simulations. These images\footnote{All images are available at \url{https://doi.org/10.6084/m9.figshare.7998143} and \url{https://doi.org/10.6084/m9.figshare.7998134}} are shown on \Fig{fig:imageset}. Three of these  were built using very large  images eventually padded with mirror boundaries to the size $20160\times20160$ pixels. One pure phase and one pure transmittance object were built from vector graphic defined USAF-1951 resolution target discretized at the right resolution.   \changes{The propagation is perform with pixels of size   $104\,\textrm{nm} < \lambda/4$ to ensure that no aliasing occurs in the intensity modeling}.
From the object plane to the detector plane, the propagation was modeled using BLAS (Band-Limited Angular Spectrum) model \cite{Matsushima2009} that implements a version of the angular spectrum propagation kernel filtered to prevent aliasing. As all the propagations are performed over the same large \fov and with the same small pixel-size, this filtering has no influence on our estimations. 
The physical parameters used all experiments are given in Tab.\ref{tab:parameters}. All  numerical experiments were   implemented\footnote{code available on \url{https://github.com/FerreolS/COMCI}}  within  the framework of the GlobalBioIm library \cite{Soubies2019}.

\begin{table}[ht!] 
    \centering
    \begin{tabular}{cccc}
        $\lambda$ & $n$ & $z$ & pixel size \\
        \hline
        $530\,$nm & 1 &  $250\,\micron$ & $104\,\textrm{nm}$\\
    \end{tabular}
        \caption{\label{tab:parameters} Parameters used in all simulations}
\end{table}

\subsection{Measurements noise}
To assess the quality of the bounds derived in \Sec{sec:fov} for \emph{i.i.d.} Gaussian noise, the modeling error $E(t)$ is estimated for two cases: (i) for a fixed noise level and varying the incidence angle, (ii) for a fixed incidence angle and varying the noise level. All the curves are computed by averaging the error over 100 different simulations.

\begin{figure}\centering
  \includegraphics[height=0.5\linewidth]{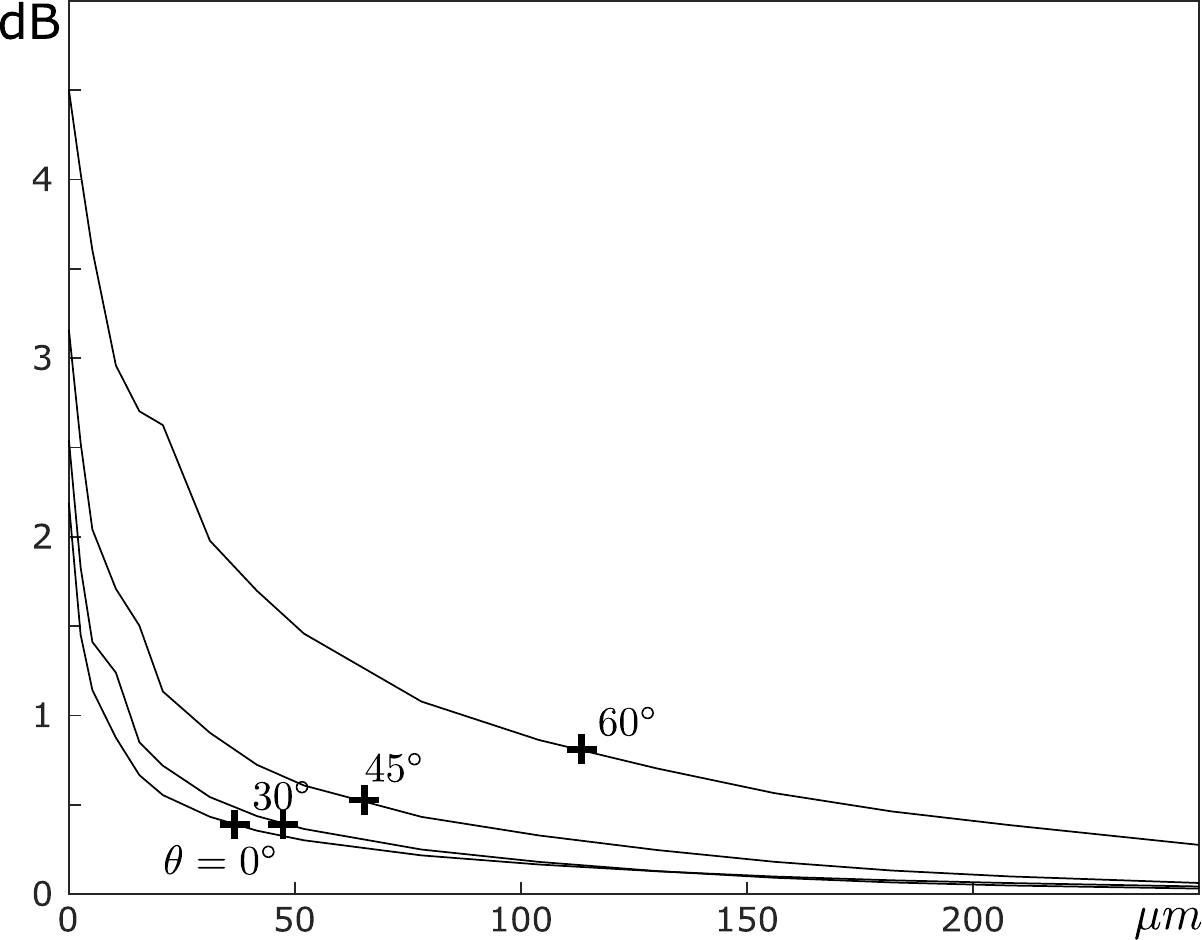}
  \caption{Modeling error $E$ as a function of the   projected distance $t$ of the opaque screen, for incidence $[0^\circ,30^\circ,45^\circ,60^\circ]$ and a noise level of $20\,$dB. The bounds given by \Eq{eq:NoiseBnd}  ($[37, 47, 75, 113]\,\micron$ respectively) are indicated by the cross marks.}
  \label{fig:SNRNoise}
\end{figure}

\begin{figure}\centering
  \includegraphics[height=0.5\linewidth]{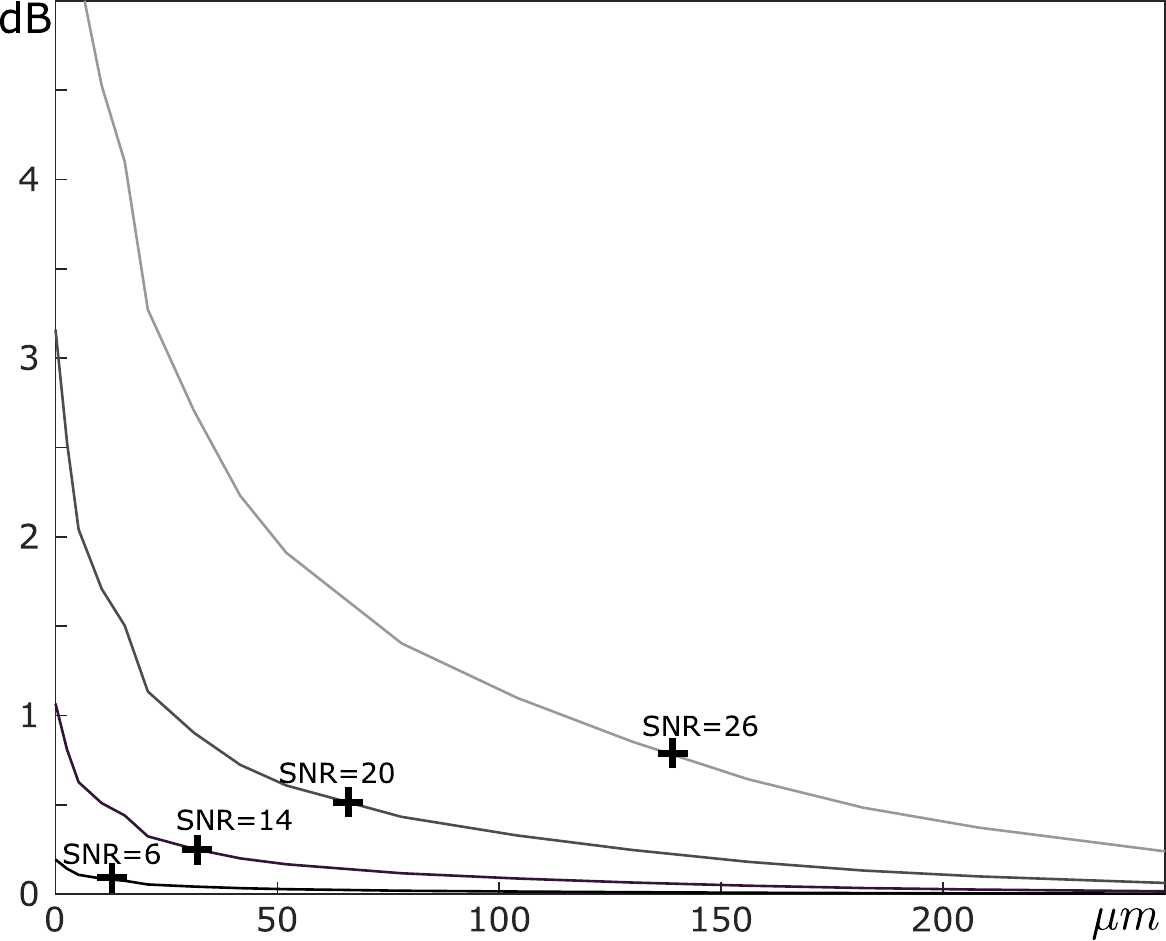}
  \caption{Modeling error $E$ as a function of the projected distance $t$  of the opaque screen for an incidence of $\theta=45^\circ$ and noise levels of $[6, 14, 20, 26]\,$dBs. The bounds given by \Eq{eq:NoiseBnd}  ($[12, 32, 65, 139]\,\micron$ respectively) are indicated by the cross marks.}
  \label{fig:NoiseIncidence}
\end{figure}

Figure \ref{fig:SNRNoise} shows the modeling error for incidence angles of  $\theta_1= [0^\circ,30^\circ,45^\circ,60^\circ]$  and a noise level of $20\,$dB. As the width of the diffraction pattern is independent of the sign of the incidence angle, negative incidence would have produces similar curves.
 Figure \ref{fig:NoiseIncidence} shows the modeling error for an incidence angle of $\theta_1 = 45^\circ$ and noise level of  $[6, 14, 20, 26] $\,dB.
 The bounds derived in \Eq{eq:NoiseBnd} are plotted on these figures. They are approximately at the bend of each plot and  provide a good trade-off between the size of the \fov and the modeling error.

\subsection{Temporal Coherence}
\begin{figure}\centering
  \includegraphics[height=0.5\linewidth]{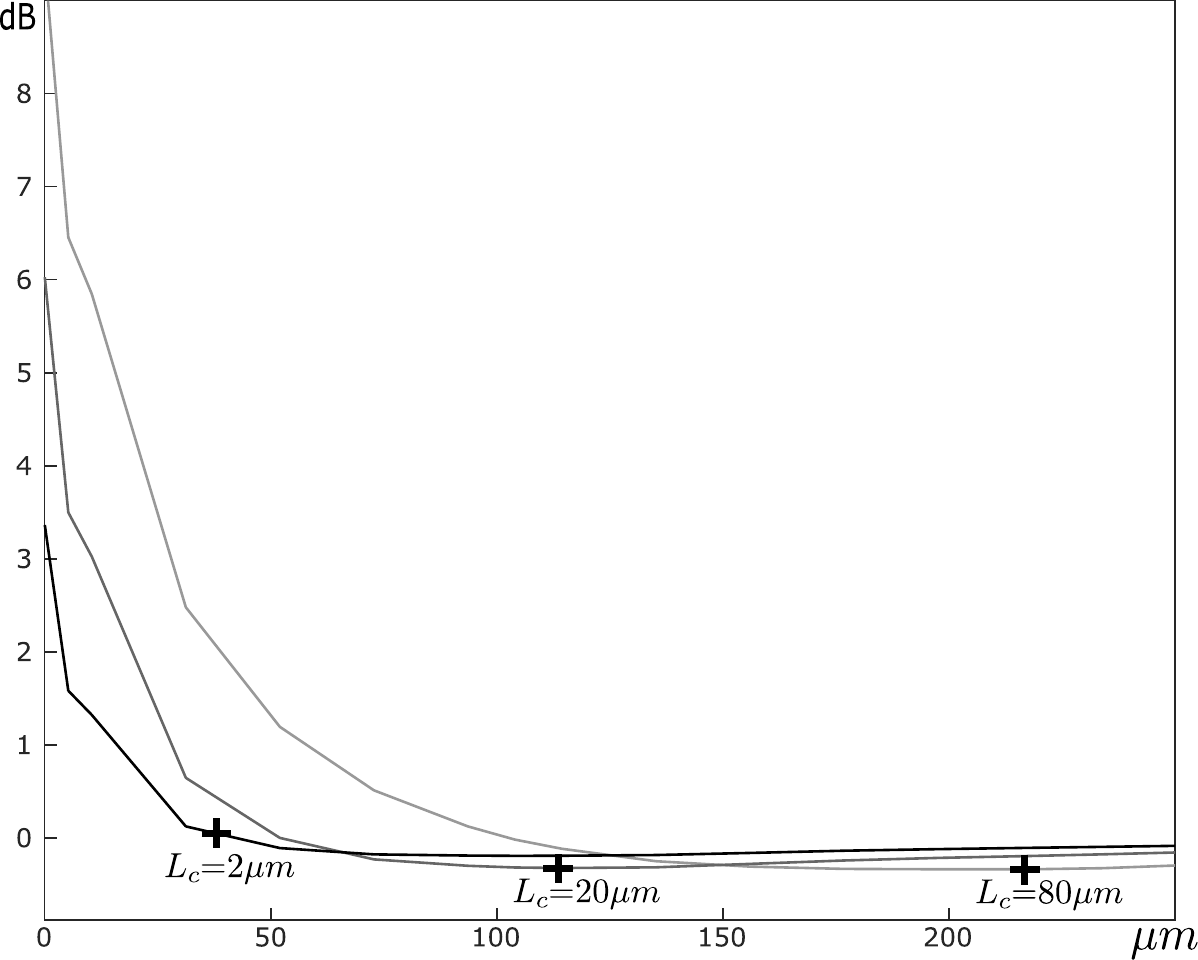}
  \caption{Modeling error $E(t)$  as a function of the projected distance $t$  of the opaque screen for an incidence  $\theta=-30^\circ$ and  coherence lengths of the illumination $L_c = [2, 20, 80]\,\micron$. The bounds given by \Eq{eq:SpeCohe} ($[38, 113, 217]\,\micron$ respectively) are indicated by the cross marks.}
  \label{fig:SNRTemporal}
\end{figure}

 The effect of a finite temporal coherence on  the  reference $\V{r}$ is computed using Monte-Carlo method averaging 100  intensities in the sensor plane for wavelengths drawn under a Normal law  modeling an illumination source with a Gaussian emission spectrum. Figure \ref{fig:SNRTemporal} shows the modeling error $E(t)$  for an incidence of  $\theta_1=-30^\circ$ and   several coherence lengths $L_c = [2, 20, 80]\,\micron$ typical of light sources used for digital holography (from white light to laser diode).

\subsection{Spatial coherence}
\begin{figure}\centering
  \includegraphics[height=0.5\linewidth]{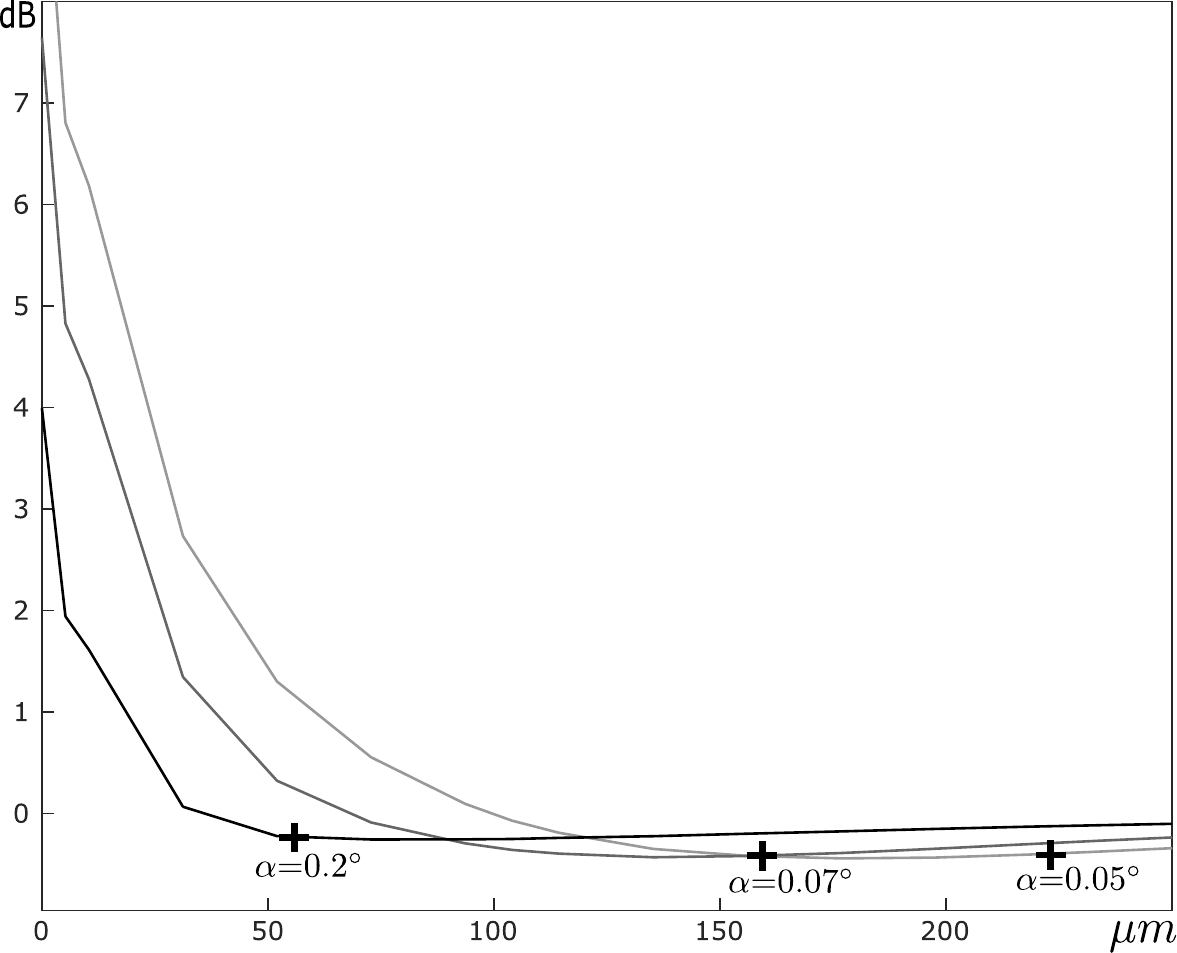}
  \caption{Modeling error $E(t)$  as a function of the projected distance $t$  of the opaque screen for an incidence  $\theta=-30^\circ$ and several angular radius  $\alpha= [0.05^\circ, 0.07^\circ, 0.2^\circ]$. The bounds given by \Eq{eq:SpaCohe} ($[56, 160, 223]\,\micron$ respectively) are indicated by the cross marks.}
  \label{fig:SNRSpat}
\end{figure}

 We simulate the reference $\V{r}$ as the hologram of the sample illuminated by an uniformly bright incoherent quasi-monochromatic circular source. To that end, we  model  the  effect of a finite  spatial coherence using Monte-Carlo method averaging 100  intensities for illumination incidence drawn uniformly within the angular size of the source. Figure \ref{fig:SNRSpat} shows the modeling error for a source placed at  $\theta_1=-30^\circ$ and with different angular radius $\alpha= [0.05^\circ, 0.07^\circ, 0.2^\circ]$.

\section{Concluding remarks}

Due to the band-limited nature of light propagation, the diffraction patterns are  theoretically unlimited in space. Hence the theoretical \fov of a lensless holographic setup is overwhelmingly extended. Sampled at $\lambda/2$, this lead to an unthinkable theoretical number of pixels needed to model perfectly the light in the detector plane. 

To model precisely enough the light propagation   while keeping the number of pixels needed acceptable, this manuscript provides theoretically grounded estimates of the \fov size and the bandwidth of a lensless holographic setup. The derived bounds are easy to estimate in practice as they depend on usually known parameters of the setup, namely the noise and the quantization level of the sensor (Eq.\ref{eq:NoiseBnd}), the coherence length (Eq.\ref{eq:SpeCohe}) and the coherence area  (Eq.\ref{eq:SpaCohe})  of the illumination source. From the size of \fov given by these bounds, we derive its bandwidth and a bound on the pixel size (Eq.\ref{eq:pixelpitch}) needed to prevent aliasing in the modeling. 

\changes{In addition to the numerical assessment of the quality of the derived bounds in the previous section, we also compute them for several already published studies with varying experimental parameters. We compare the claimed \fov{} and/or resolution with the bounds derived in this paper::
\begin{itemize}
    \item In \cite{Soulez2007}, particles were detected over a \fov{} of $34\times 27\,\textrm{mm}^2$, 4 times larger than the sensor area $8.6\times6.9\,\textrm{mm}^2$. Supposing a realistic SNR of $40\,$dB, the experimental conditions ($\lambda=523\,$nm, $z=250\,$mm and $\theta=0$)  lead to a half width of the diffraction pattern of $p_i^{\RM{noise}}= 12$mm giving a similar \fov{} of   $33\times 31\,\textrm{mm}^2$.
    \item In \cite{Fournier2016}, the hologram is reconstructed over a \fov{} of $23\times 23\,\textrm{mm}^2$, 3 times larger than the sensor area $7.6\times7.6\,\textrm{mm}^2$. With the  experimental conditions  ($\lambda=662\,$nm, $z=283\,$mm and $\theta=0$), a SNR of $35\,$dB leads to a half width of the diffraction pattern of $p_i^{\RM{noise}}= 18\,$mm giving a similar \fov{} of  $23\times 23\,\textrm{mm}^2$. 
    \item In \cite{Luo2015}, whereas no \fov extrapolation is performed, the resolution claimed is $250\,$nm under a highly inclined illumination with a partially coherent source of  coherence length $L_c=62\,$nm.  The  experimental conditions  ($\lambda=700\,$nm, $z=0.1\,$mm, $n=1.52$ and $\max(\theta)=52^\circ$) lead to a half width of the diffraction given by \Eq{eq:SpeCohe} of $p_i^{\RM{spe}}= 7.7\,$mm and a resolution  as given by \Eq{eq:pixelpitch} of $R= 259\,$nm close to the claimed resolution of $250\,$nm. This slight difference in resolution can be explained by the small spatial support of the grating lines observed that sufficiently broaden its spatial frequencies to fit within the setup spatial bandwidth\cite{Kelly1965}.
\end{itemize}}

This work have been done in the context of 2D imaging of thin sample. However its extension in 3D is straightforward as the capability to distinguish two different planes in a lensless setup can be deduced from its bandwidth as shown in \Sec{sec:bandwidth}. Depending on the application,  larger \fov and bandwidth are not always  better and the proposed bounds can be used to design the coherence of the illumination 
to reduce cross interference between scatterers as in \cite{Mudanyali2010}.

\section{Acknowledgements and funding}
 This work was supported by Auvergne Rhône Alpes Region through the DIAGHOLO project. The author thank A. Berdeu, E. Thiébaut and M. Tallon for their careful readings and numerous suggestions to improve this paper.\\
 \textbf{Disclosures.} The author declares no conflicts of interest.
\bibliography{BiblioPhase,dataset}

\bibliographystyle{unsrt}

\end{document}